\documentclass[twocolumn,aps,amsmath,amssymb,floatfix,prl,superscriptaddress,nofootinbib]{revtex4}
\usepackage{bbm}
\usepackage[dvips]{graphicx}
\usepackage{amsmath,amssymb,latexsym}

\usepackage{xcolor}
\renewcommand{\ol}[1]{\overline{#1}}
\renewcommand{\L}{\mathcal{L}}
\newcommand{\N}{\mathcal{N}}
\newcommand{\R}{\mathcal{R}}
\renewcommand{\Re}{\mathrm{Re}\,}

\newcommand{\e}{\mathbf{e}}
\newcommand{\f}{\mathbf{f}}
\renewcommand{\k}{\mathbf{k}}
\newcommand{\ki}{{\mathbf{k}_{\mathrm{I}}}}
\newcommand{\kii}{{\mathbf{k}_{\mathrm{II}}}}
\renewcommand{\l}{\mathbf{m}} %

\renewcommand{\u}{\mathbf{u}}
\newcommand{\x}{\mathbf{x}}
\newcommand{\Mmat}{\mathbf{M}}
\newcommand{\Q}{\mathbf{Q}}
\newcommand{\micron}{\mu\mathrm{m}} %
\newcommand{\comma}{\quad,}

\newcommand{\zero}{{\boldsymbol{0}}}
\newcommand{\K}{\mathbbm{K}}

\renewcommand*{\vec}[1]{\mathbf{#1}}

\newcommand*{\vf}{\varphi}
\newcommand*{\F}{\boldsymbol{\Phi}}
\newcommand*{\G}{\Gamma}
\newcommand*{\Gmat}{\boldsymbol{\Gamma}} %
\newcommand{\gp}{\overline{\varphi}} %
\newcommand*{\FP}{\boldsymbol{\Phi}^\ast}%
\newcommand*{\Lmap}{\mathcal{L}} %
\newcommand*{\Lmat}{\mathbf{L}}  %
\newcommand*{\D}{\boldsymbol{\Delta}} %

\begin{document}

\title{Synchronization in cilia carpets:\\ multiple metachronal waves are stable, but one wave dominates}
\author{Anton Solovev}
\affiliation{TU Dresden, Dresden, Germany}
\author{Benjamin M. Friedrich}
\email{benjamin.m.friedrich@tu-dresden.de}
\affiliation{TU Dresden, Dresden, Germany}

\date{\today}

\begin{abstract}
Carpets of actively bending cilia represent arrays of biological oscillators that can exhibit self-organized metachronal synchronization
in the form of traveling waves of cilia phase.
This metachronal coordination supposedly enhances fluid transport by cilia carpets.
Using a multi-scale model calibrated by an experimental cilia beat pattern,
we predict multi-stability of wave modes.
Yet, a single mode, corresponding to a dexioplectic wave, has predominant basin-of-attraction.
Similar to a ``dynamic'' Mermin-Wagner theorem,
relaxation times diverge with system size, which rules out global order in infinite systems.
In finite systems, we characterize the synchronization transition as function of quenched frequency disorder,
using generalized Kuramoto order parameters.
Our framework termed \textit{Lagrangian Mechanics of Active Systems} 
allows to predict the direction and stability of metachronal synchronization for given beat patterns.
\end{abstract}

\maketitle

Motile cilia are slender cell appendages that bend rhythmically due to the activity of molecular dynein motors inside~\cite{Gray1928}.
Collections of motile cilia can spontaneously synchronize their bending waves,
e.g., in carpets of many cilia on airway epithelium~\cite{Sanderson1981},
as well as on the surface of model organisms, e.g., green alga colonies or unicellular \textit{Paramecium}~\cite{Machemer1972,Brumley2012}.
Metachronal coordination manifests itself as a self-organized traveling wave of cilia phase
(similar to a Mexican wave in a soccer stadium).
Numerical models showed that this synchronization is important for efficient fluid transport~\cite{Osterman2011,Elgeti2013}.
Tissue-scale polarity systems align cilia bases~\cite{Guirao2010},
ensuring a common direction of the effective stroke of the cilia beat.
In many species, cilia beat patterns are chiral,
e.g., with counter-clockwise motion of cilia during their recovery stroke close to the surface~\cite{Machemer1972}.
The directions of metachronal waves enclose defined angles
relative to the direction of the effective stroke~\cite{Machemer1972,Knight1954},
presumably set by the chirality of the cilia beat~\cite{Meng2021}.

Already in 1952, Taylor proposed that hydrodynamic interactions between nearby cilia play a key role for their synchronization~\cite{Taylor1952}.
When a beating cilium performs its bending wave,
it sets the surrounding fluid in motion,
resulting in time-dependent hydrodynamic friction forces that act on nearby cilia.
Recent experiments indeed demonstrated synchronization by hydrodynamic coupling in pairs of cilia~\cite{Brumley2014},
as well as phase-locking to external oscillatory flows
with characteristic Arnold tongues
\cite{Quaranta2015,Pellicciotta2020}.
Recent theoretical work predicts different synchronization modes between pairs of hydrodynamically coupled cilia, 
depending on their relative positions~\cite{Solovev2020a, Man2020}.

Yet, we still do not understand how hydrodynamic interactions and the shape of the cilia beat
select the direction of metachronal waves in cilia carpets.
Multiple wave directions are possible,
yet these may not be stable to small perturbations (\textit{local stability})
or be unlikely to be selected for random initial conditions (\textit{global stability}).
A key question thus concerns the local and global stability of different metachronal wave modes.
The global stability of synchronization states in collections of coupled oscillators,
not just interacting cilia, is still a field of active research~\cite{Wiley2006, Delabays2017, Menck2013}.

The periodic sequence of shapes that a cilium assumes during its beat cycle represents a limit cycle
in a high-dimensional shape space~\cite{Ma2014,Werner2014}.
This limit cycle can be parameterized by a single phase variable
such that phase speed is constant in the absence of perturbations and noise~\cite{Pikovsky2003}.
This allows to describe beating cilia as phase oscillators~\cite{Ma2014, Wan2014b}.
In the presence of external flows, which change the hydrodynamic load,
the phase speed changes, i.e.,
cilia progress slower or faster along their beat cycle,
while deviations from the limit-cycle sequence of shapes remain small for moderate flows
\cite{Goldstein2009,Klindt2016,Pellicciotta2020}.
This \textit{load-response} of cilia
(reflected by the load-dependent speed of their phase variable)
is a prerequisite for cilia synchronization by hydrodynamic interactions,
and is implicit in previous minimal models
\cite{Vilfan2006,Guirao2007,Niedermayer2008,Uchida2011,Wollin2011,Friedrich2012,Friedrich2016,Pellicciotta2020,Meng2021}.

Previous theory on hydrodynamic synchronization in cilia carpets
either employed large-scale numerical simulations~\cite{Gueron1999,Elgeti2013,Stein2019}.
or relied on minimal models, where beating cilia are idealized, e.g., as orbiting spheres
\cite{Vilfan2006,Guirao2007,Niedermayer2008,Uchida2011,Wollin2011,Friedrich2012,Friedrich2016,Pellicciotta2020,Meng2021}.

Here, we harness multi-scale simulations to
combine the benefits of detailed hydrodynamic simulations based on experimentally measured cilia beat patterns, and
those of minimal models amenable to local and global stability analysis.
Our approach, termed \textit{Lagrangian mechanics of active systems}~\cite{Solovev2020a},
enables us to study global stability in arrays of hydrodynamically coupled cilia.

 \paragraph{Beating cilia as coupled phase oscillators.}

\begin{figure*}
\includegraphics[width=0.8\linewidth]{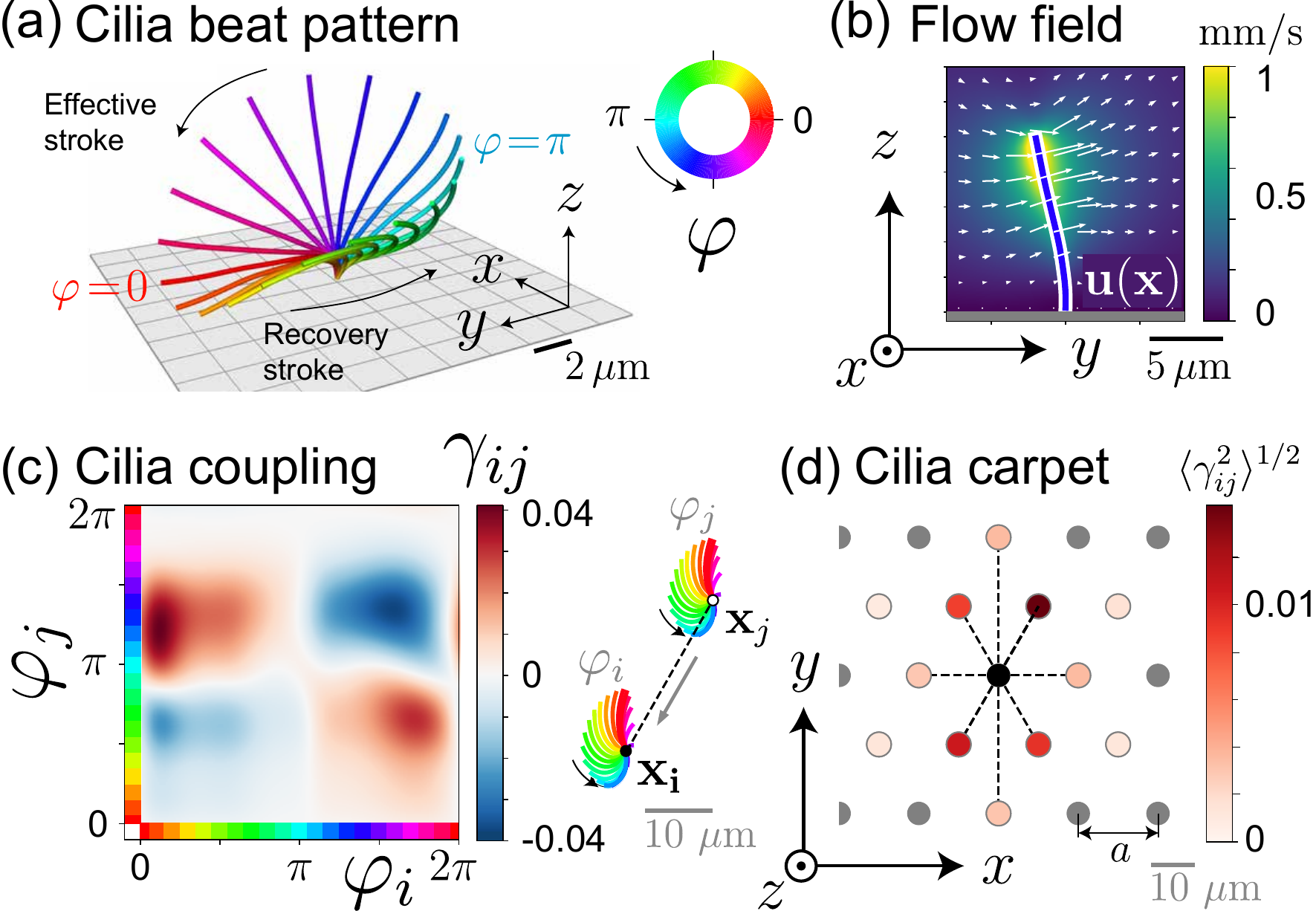}
\caption{
\textbf{Multi-scale model of hydrodynamic synchronization in cilia carpets.}
(a) Cilia beat pattern from~\cite{Machemer1972,Naitoh1984},
whose periodic shape sequence has been parameterized by a $2\pi$-periodic phase variable $\varphi$ (color code).
(b) Computed flow field $\u$ for this beat pattern
(colors: $|\u(\x)|$, arrows: projection of $\u$ on $yz$-plane).
(c) Normalized hydrodynamic interaction
$\gamma_{ij}(\varphi_i, \varphi_j)=\Gamma_{ij}(\varphi_i, \varphi_j)/\Gamma_{ii}(\varphi_i)$
between a pair of cilia
as function of their phases $\varphi_i$ and $\varphi_j$:
Positive values $\gamma_{ij}$ cause cilium $i$ to beat slower, see Eq.~(\ref{eq:motion}).
Separation vector of cilia bases,
$\x_j-\x_i=a\,(\cos\psi\,\e_x + \sin\psi\,\e_y)$, $\psi{=}\pi/3$.
(d) Triangular lattice of cilia base points $\x_j$ (dots).
The color-code represents the root-mean-square average $\langle \gamma_{ij}^2 \rangle^{1/2}$
of the normalized hydrodynamic interaction $\gamma_{ij}$
between the cilium with base $\x_j$ (colored dot) and a central cilium at $\x_i$ (black dot).
Dashed lines indicate hydrodynamic interactions included in our cilia carpet model.
Lattice spacing $a=18\,\micron$,
intrinsic cilium beat frequency $\omega_0 /(2\pi) = 32 \,\mathrm{Hz}$~\cite{Machemer1972};
panel (b):
$\varphi=1.4\pi$.
}
\label{figure1}
\end{figure*}

We consider a carpet of $N$ cilia positioned on a regular triangular lattice of base points $\x_j$
in a rectangular domain with periodic boundary conditions,
see Fig.~\ref{figure1}(d).
Each cilium is described as a phase oscillator whose phase $\varphi_j$ advances by $2\pi$ on each cycle, like a clock.
This phase variable $\varphi_j$ parameterizes a periodic sequence of three-dimensional cilia shapes,
previously measured for \textit{Paramecium}~\cite{Machemer1972,Naitoh1984}, see Fig.~\ref{figure1}(a).
When the phase $\varphi_j$ increases, i.e., the cilium progresses along its beat cycle,
the corresponding shape change of the cilium sets the surrounding fluid in motion,
resulting in time-dependent hydrodynamic friction forces that act on the other cilia.
For nearby cilia,
the resultant hydrodynamic interactions
can be computed from the Stokes equation valid at zero Reynolds number~\cite{Wei2019,Solovev2020a},
see also Supplemental Material (SM).
The plane containing the cilia base points is modeled as a non-slip boundary,
thus hydrodynamic interactions decay as $1/d^3$ as function of distance $d$~\cite{Blake1974a,Solovev2020a}.

We consider the dynamics of $N$ cilia in a rectangular unit cell with periodic boundary conditions,
which is characterized by a $N$-component vector
$\F = (\varphi_1,\ldots,\varphi_N)\in\mathbb{R}^N$
of cilia phases.
Because the Stokes equation is linear~\cite{Happel:hydro},
the surface density of hydrodynamic friction forces $\f(\x)$ at time $t$
(defined on the combined surface $\mathcal{S}$ of all cilia and the boundary surface)
is linear in the generalized velocity $\dot{\F}$.
Thus, the power exerted by the moving cilia on the surrounding fluid
$\R=\int_\mathcal{S} d^2\x\, \f(\x)\cdot \dot{\x}$
becomes a quadratic form in $\dot{\F}$~\cite{Solovev2020a}
\begin{equation}
\label{eq:R}
\R = \dot{\F} \cdot \boldsymbol{\Gamma}(\F) \cdot \dot{\F}
\end{equation}
with a symmetric $N\times N$ matrix of generalized hydrodynamic friction coefficients
$\boldsymbol{\Gamma}=\boldsymbol{\Gamma}(\F)$.
Here,
$\Gamma_{ii}$ represents self-friction of cilium~$i$,
while $\Gamma_{ij}$ characterizes hydrodynamic interactions between cilia $i$ and $j$.
Below, we compute $\boldsymbol{\Gamma}(\F)$ in a pairwise-interaction approximation.

For each cilium, we introduce the generalized hydrodynamic friction force $P_i$
as the friction force conjugate to the generalized coordinate $\varphi_i$
(following the formalism of Lagrangian mechanics of dissipative systems
with $\R/2$ as Rayleigh dissipation function~\cite{Goldstein:classical_mechanics,Solovev2020a})
\begin{equation}
\label{eq:P}
P_i=  \frac{1}{2} \partial \mathcal{R}/\partial \dot{\varphi}_i  = \ \sum_j \Gamma_{ij} \dot{\varphi}_j
\quad.
\end{equation}
Assuming low Reynolds numbers,
there is at all times a \textit{force balance}
between the generalized friction force $P_i$ and 
an active driving force $Q_i$ that coarse-grains the active processes inside cilium $i$ that drive the cilia beat
\begin{equation}
\label{eq:force_balance}
Q_i(\varphi_i) = P_i(\F,\dot{\F}) \quad,\quad i=1,\ldots,N \quad.
\end{equation}
The cilia driving force $Q_i$ is an intrinsic property of cilium $i$,
hence only depends on $\varphi_i$, and possibly load $P_i$.
We make the simplifying assumption that $Q_i$ is independent of load.
Previous experiments in the green alga \textit{Chlamydomonas}~\cite{Klindt2016}
as well as cilia bundles in external flow~\cite{Pellicciotta2020}
showed that this assumption together with Eq.~\eqref{eq:force_balance}
quantitatively accounts for the \textit{load response} of cilia~\cite{Friedrich2016,Friedrich2018},
i.e., the experimental observation that cilia progress slower/faster along their beat cycle upon increase/decrease of hydrodynamic load.
Next, we compute the generalized hydrodynamic friction forces $P_i{=}\sum_j \Gamma_{ij}\dot{\varphi}_j$
with friction coefficients $\Gamma_{ij}$ for a real cilia beat pattern,
and calibrate the active driving forces $Q_i$.
Note that previous minimal models of hydrodynamically interacting spheres
\cite{Vilfan2006,Guirao2007,Niedermayer2008,Uchida2011,Wollin2011,Friedrich2012}
can likewise be written in the form of Eq.~\eqref{eq:force_balance}, yet with simplified driving and friction forces.

\paragraph{Oscillator coupling calibrated from hydrodynamic simulations.}

Initial simulations showed that the friction coefficient
$\Gamma_{ij}(\F)$
is largely independent of the phases of the other cilia,
$\varphi_k$, $k\neq i,j$.
This allows us to use an approximation of only pairwise-interactions for $\Gmat(\F)$
by averaging out all non-essential variables.
In short, we set $\varphi_k=\varphi$ for $k\neq i,j$,
and average over $\varphi$ to obtain a function $\Gamma_{ij}(\varphi_i,\varphi_j)$
of $\varphi_i$ and $\varphi_j$ only, see SM text for details.
The active driving force $Q_i(\varphi_i)$ of each cilium
is uniquely determined by a reference condition,
namely that the phase speed of this cilium should be constant, $\dot{\varphi}_i = \omega_0$,
if the other cilia do not beat.
This condition yields
\begin{equation}
\label{eq:Q}
Q_i(\varphi_i)=\omega_0\,\Gamma_{ii}(\varphi_i)
\quad.
\end{equation}
Together, Eqs.~(\ref{eq:P}), (\ref{eq:force_balance}) and (\ref{eq:Q})
give an \textit{equation of motion} in implicit form
\begin{equation}
\dot{\varphi}_i = \omega_0 - \sum_{j\neq i} \gamma_{ij}\, \dot{\varphi}_j
\text{ with }
\gamma_{ij}(\varphi_i,\varphi_j) = \frac{\Gamma_{ij}(\varphi_i,\varphi_j)}{\Gamma_{ii}(\varphi_i)}
\quad.
\label{eq:motion}
\end{equation}
The normalized hydrodynamic interaction
$\gamma_{ij}(\varphi_i,\varphi_j)$
between cilium $i$ and cilium $j$
characterizes the relative amount by which the motion of cilium $j$ changes the phase speed of cilium $i$.
Fig.~\ref{figure1}(c) shows
$\gamma_{ij}(\varphi_i,\varphi_j)$
as function of the respective phases $\varphi_i$ and $\varphi_j$ of the two cilia.
In short, the effective stroke of cilium $j$ ($\pi \lesssim \varphi_j \lesssim 2\pi$)
will speed up cilium $i$ ($\gamma_{ij}{<}0$, blue colors) if cilium $i$ is also in its effective stroke ($\pi \lesssim \varphi_i \lesssim 2\pi$),
but will slow down cilium $i$ ($\gamma_{ij}{>}0$, red colors) if cilium $i$ is in its recovery stroke ($0 \lesssim \varphi_j \lesssim \pi$).
When one of the two cilia transitions from effective stroke to recovery stroke, 
or vice versa (i.e., $\varphi_i\approx 0$, $\pi$ or $\varphi_j\approx 0$, $\pi$), 
that cilium moves slowly and the hydrodynamic interaction between the two cilia is weak, 
$\gamma_{ij}\approx 0$.
We emphasize that $\gamma_{ij}(\varphi_i,\varphi_j)$ is not simply a function of the phase difference $\varphi_i-\varphi_j$ 
as in a classical Kuramoto model, but is much richer.

Numerical computations further show that $\gamma_{ij}$ is very small except for close neighbors;
we therefore set $\gamma_{ij}=0$ except for close neighbors $i$ and $j$, see Fig.~\ref{figure1}(d).
We can now rewrite the equation of motion equivalently in explicit form as
$\dot{\F} = \Gmat^{-1}\cdot\Q$.
With pre-computed $\Gamma_{ij}(\varphi_i,\varphi_j)$ and $Q_i(\varphi_i)$ at hand,
this explicit ordinary differential equation can be efficiently integrated for ten-thousands of cilia beat cycles.

\begin{figure*}
\includegraphics[width=\linewidth]{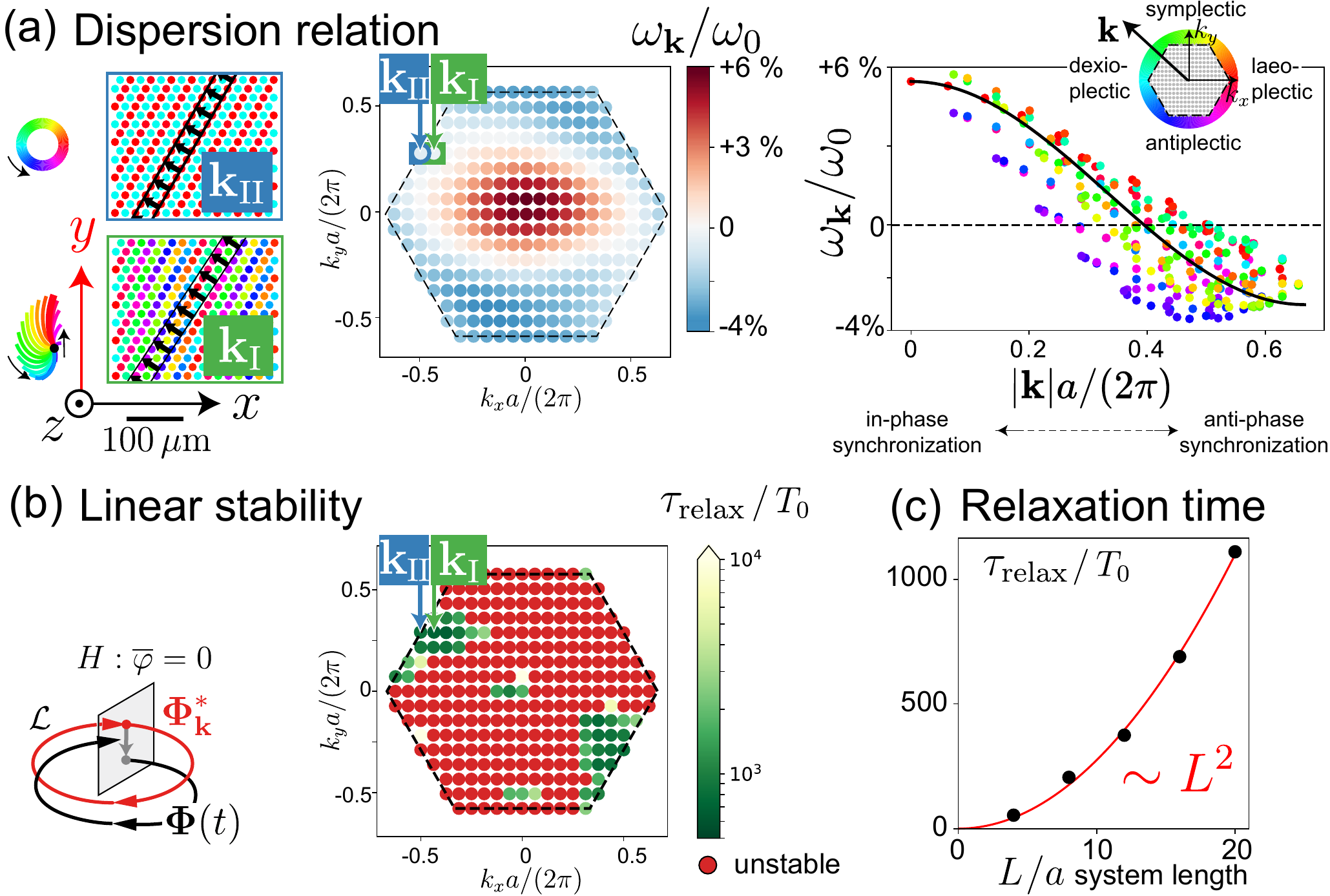}
\caption{
\textbf{Multi-stability of metachronal waves.}
(a)
\textit{Dispersion relation:}
Left:
Two example metachronal wave solutions:
colored dots mark cilia base points, with colors representing cilia phase at a snapshot in time.
Middle:
Metachronal wave solutions can be enumerated by a finite set of $N$ wave vectors $\k$
in a Brillouin zone (colored dots, example wave vectors $\k_\mathrm{I}$, $\k_\mathrm{II}$ highlighted).
Colors represent the angular frequency $\omega_\k$ of wave solutions
(normalized by the intrinsic frequency $\omega_0$ of a single cilium).
Right: Wave frequency $\omega_\k/\omega_0$ re-plotted as function of inverse wavelength
$|\k|a/(2\pi)$:
cilia beat faster for long-wavelength coordination ($|\k|$ small, approaching \textit{in-phase} synchronization)
as compared to metachronal coordination with short wavelength ($|\k|$ large, approaching \textit{anti-phase} synchronization).
The wave frequencies approximately follow an analytical result
$\Delta \omega_\k \sim \cos(\pi|\k|/k_\mathrm{max})$
for a classical Kuramoto model (black line).
Different colors indicate the direction of $\k$, see inset.
Traditionally, wave directions are classified as \textit{symplectic}, \textit{antiplectic}, \textit{dexioplectic}, \textit{laeoplectic},
depending on the direction of $\k$ relative to the direction $\e_y$ of the cilia effective stroke~\cite{Knight1954}.
(b)
\textit{Linear stability:}
Linear stability analysis for each $\k$ reveals that multiple solutions are linearly stable
(green colors:
stable metachronal wave solution,
color represents relaxation time $\tau_\text{relax}$ of the slowest decaying perturbation mode,
normalized by beat period $T_0=2\pi/\omega_0$ of single cilium;
red: unstable).
For the computation,
we define a global phase $\ol{\varphi}$ and
analyze the stroboscopic dynamics of the cilia carpet given by $\ol{\varphi}=0$ modulo $2\pi$:
fixed points $\FP_\k$ of this Poincar{\'e} map correspond to
metachronal wave solutions, see left inset.
(c)
The relaxation time of the slowest-decaying perturbation for the dominant wave solution
increases with system length as $\sim L^2$,
resembling a dynamic Mermin-Wagner theorem for cilia carpets,
which rules out global order in infinite systems.
Lattice of $16\times 16$ cilia; other parameters as in Fig.~\ref{figure1}. 
}
\label{figure2}
\end{figure*}

\paragraph{Metachronal wave solutions.}
We are interested in dynamic steady-state solutions of the equation of motion, Eq.~(\ref{eq:motion}).
As a reference, we first re-visit the classical Kuramoto model with local sinusoidal coupling~\cite{Sarkar2021,Doerfler2014}
specifically, we consider a Kuramoto model of coupled phase oscillators
with phases $\varphi_i$ at respective lattice positions $\x_i$
and equation of motion
$
\dot \varphi_j(t)  = \omega_0 - \sum_{i\neq j} c_{ij}(\varphi_i, \varphi_j)
$
with coupling function
$c_{ij}=\varepsilon\sin(\varphi_j-\varphi_i)$
for all pairs $(i,j)$ of neighbors and $c_{ij}=0$ else.
For this Kuramoto model,
the steady-state solution are perfect plane traveling waves 
with wave vector $\k$
\begin{equation}
\label{eq:wave}
\F_\k(t): \varphi_j(t) = \omega_\k\, t - \k \cdot \x_j \quad.
\end{equation}
Here, $\vec{k}$ is one of the $N$ reciprocal lattice points in the Brillouin zone of the oscillator lattice
(with unit cell of $N$ oscillators and periodic boundary conditions),
see also Fig.~\ref{figure2}(a). 
Note $\omega_\k=\omega_0$ for this simple Kuramoto model.
In our cilia carpet model,
the hydrodynamic interaction coefficients $\gamma_{ij}$ are not perfect sinusoidal functions, but a superposition of many Fourier modes.
As a consequence, periodic solutions of cilia carpet dynamics are not perfect plane traveling waves 
as in Eq.~(\ref{eq:wave}).
Nonetheless, we numerically find $N$ periodic wave solutions $\FP_\vec{k}(t)$
of cilia carpet dynamics,
where each $\FP_\vec{k}(t)$ is close to one of the $N$ plane traveling wave $\F_\vec{k}(t)$ of Eq.~(\ref{eq:wave}).
We will refer to $\FP_\vec{k}(t)$ as \textit{metachronal wave solutions}.
The global frequency $\omega_\k$ of these periodic solutions decreases with inverse wavelength $|\k|$,
see Fig.~\ref{figure2}(a).
The numerical dispersion relation is well approximated
by $\omega_\k / \omega_{\k=\zero} \approx  1 + \beta [ \cos( \pi |\k| / k_{\max}) - 1]$
with $\beta \approx 0.04 $ and $k_{\max}=4 \pi / (3a)$,
inline with analytical results for a slightly more general Kuramoto model~\cite{Sakaguchi1986}
with $c_{ij}=\varepsilon\sin(\varphi_j-\varphi_i+\delta)$
involving an additional phase shift $\delta$ in the coupling, see SM text for details.

\paragraph{Linear stability analysis of metachronal wave solutions.}

To analyze the stability of metachronal wave solutions with respect to small perturbations,
we map periodic solutions  
onto fixed points of a suitable Poincar{\'e} map~\cite{Verhulst1996}.
We can then analyze the local stability of these fixed points using standard linear stability analysis.
We first define a continuous \textit{global phase}
as the mean $\ol{\varphi}(t) = \sum_j \varphi_j(t) / N$
for a continuous trajectory $\F(t)\in \mathbbm{R}^N$ in phase space.
Note that the mean of angular values can only be defined modulo $2\pi/N$;
yet this ambiguity is resolved if we define $\ol{\varphi}$ for an entire time-continuous trajectory.
We now define a Poincar{\'e} plane $H$ by 
setting this global phase to zero, $\gp=0$, and
a Poincar{\'e} return map $\Lmap: H \rightarrow H$,
corresponding to an increase of the global phase $\gp$ by $2\pi$
[i.e., a trajectory $\F(t)$ starting at $\F(0)=\F_0\in H$
intersects the shifted Poincar{\'e} plane $H+2\pi\,\vec{1}$ at
$\F_1=\Lmap(\F_0)+2\pi\,\vec{1}$],
see inset on the left in Fig.~\ref{figure2}(b).

Fixed points $\FP_\vec{k}$ of this Poincar{\'e} map with $\L(\FP_\vec{k}) = \FP_\vec{k}$
correspond to periodic orbits $\FP_\vec{k}(t)$ of the full dynamics.
To determine whether a metachronal wave solution is stable,
we linearize the Poincar{\'e} map at the corresponding fixed point
$\F_\k^\ast$
\begin{equation}
\label{eq:Lmat}
\Lmap (\FP_\vec{k}  + \boldsymbol{\Delta}) \approx
\FP_\k +  \Lmat_\k \cdot \boldsymbol{\Delta}\quad.
\end{equation}
The eigenvalues $\lambda_1,\ldots,\lambda_{N-1}$ of $\ln(\Lmat_\k)$
represent dimensionless Lyapunov exponents
(whose real parts are proportional to inverse relaxation times),
while the corresponding eigenvectors $\boldsymbol{\Delta}_1,\ldots,\boldsymbol{\Delta}_{N-1}$ represent fundamental perturbation modes.
The fixed point $\F_\k^\ast$ is linearly stable if $\mathrm{Re}\,\lambda_i<0$ for all $i$.
In this case, all perturbation modes $\boldsymbol{\Delta}_i$
decay with respective relaxation times
$\tau_i = 2\pi/|\omega_\k\,\Re\lambda_i|$.
A non-zero imaginary part of the Lyapunov exponents implies 
that perturbations decay in a spiral-like fashion to the fixed point $\F_\k^\ast$ with period
$(2\pi)^2 / |\omega_\k \, \mathrm{Im}\,\lambda_i|$.
We observe that multiple metachronal wave solutions
are simultaneously stable:
Fig.~\ref{figure2}(b)
reports the relaxation time $\tau_\mathrm{relax}=\max\tau_i$ 
of the slowest decaying perturbation mode for stable wave solutions.
The multistability of wave solutions is inline with previous observations in minimal models~\cite{Meng2021}.

\begin{figure*}
\includegraphics[width=\linewidth]{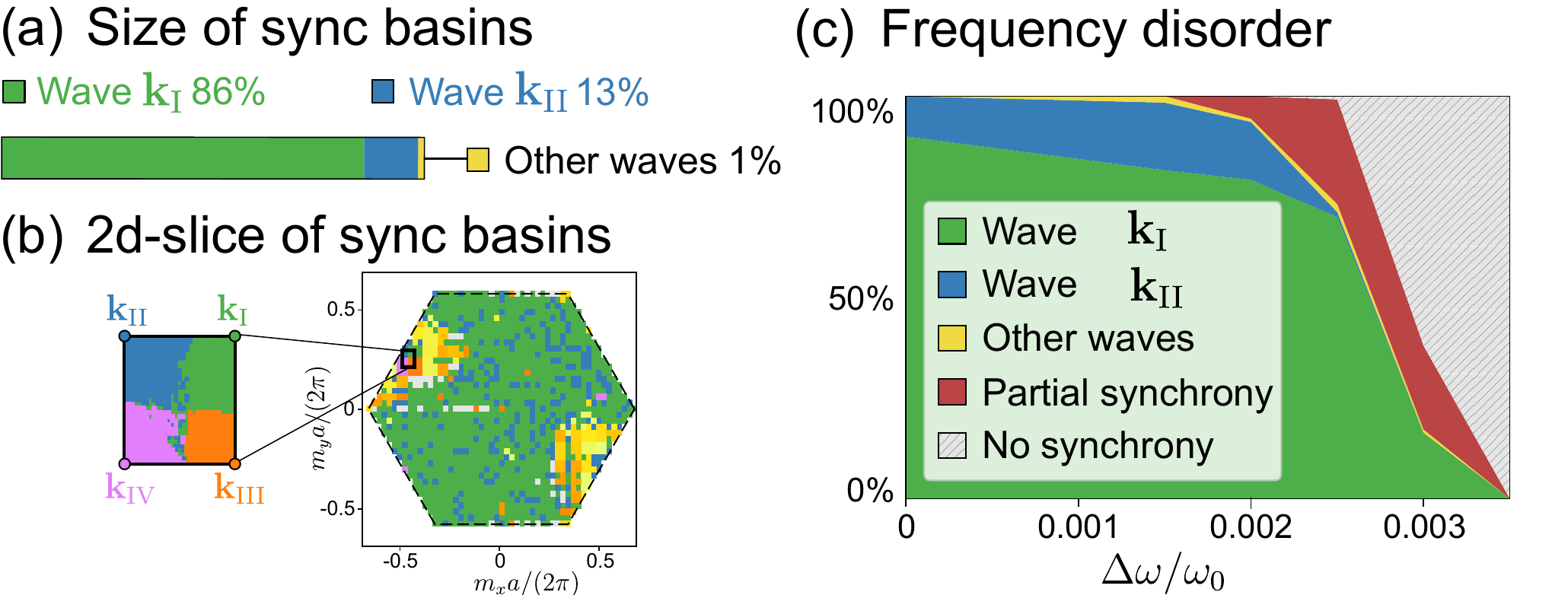} 
\caption{
\textbf{Global stability reveals dominant wave mode.}
(a)
\textit{Size of sync basins.}
We estimated the relative size of the basin-of-attractions of wave solutions $\k$ (`sync basins' \cite{Wiley2006}),
by drawing $400$ random initial conditions from a uniform distribution,
of which $86\%$ converged to one dominant wave mode $\k_\mathrm{I}$,
while $13\%$ converged to the adjacent wave mode $\k_\mathrm{II}$
[introduced in Fig.~\ref{figure2}(a)].
 (b)
\textit{Slice of sync basins.}
To visualize basins-of-attraction,
we show limit points $\FP_\k$ for special initial conditions
$\varphi_j(t{=}0) = -\x_j\cdot\l$
with off-lattice wave vector $\l$;
this choice corresponds to a two-dimensional slice through $N$-dimensional phase space.
Gray dots indicate initial conditions, for which trajectories did not converge to any $\FP_\k$.
Upon magnification, the boundaries of the basins-of-attraction appear rough, see inset to the left.
(c)
\textit{Frequency disorder.}
Relative size of basins-of-attraction for different metachronal wave modes
as function of increasing quenched disorder $\Delta\omega/\omega_0$
of intrinsic cilia beat frequencies with
$\Delta\omega^2 \approx \langle \omega_i^2 \rangle -\langle\omega_i\rangle^2$:
synchronization is lost at a characteristic disorder threshold.
For intermediate $\Delta\omega$, some realization display high order parameters $r_\k \ge 2^{-1/2}$ for some $\k$,
but not all cilia adopt a common frequency,
corresponding to a regime of \textit{partial synchronization} (red).
Parameters as in Fig.~\ref{figure2}.
}
\label{figure3}
\end{figure*}

 \paragraph{Global stability: one wave dominates.}
Although many metachronal wave solutions with different wave vectors $\k$ are simultaneously stable to small perturbations,
we find that trajectories with uniformly sampled random initial conditions
will predominantly converge to just one wave solution.
The fraction of trajectories converging to $\FP_\k$,
equals the volume fraction of the basin-of-attraction of $\FP_\k$,
which yields $86\%$
for the dominant wave solution with wave vector $\k_\mathrm{I}$, see Fig.~\ref{figure3}(a).
 
\paragraph{Slice-visualization of basins-of-attraction.}
To visualize basins-of-attractions of metachronal wave solutions, 
we additionally considered a specific set of initial conditions of the form
$\varphi_j = - \vec{m}  \cdot \x_j$ 
with ``off-lattice'' wave vectors $\vec{m}$; 
these initial conditions correspond to a two-dimensional slice through the $N$-dimensional phase space,
see Fig.~\ref{figure3}(b).
As expected, the majority of initial conditions converged to the dominant wave mode $\k_\mathrm{I}$, 
while initial conditions $\vec{m}\approx\k$ in a small neighborhood 
of other stable modes $\k$ converged to the respective $\FP_\k$.
A magnification shows that the boundaries between basins-of-attraction are rough (and possibly fractal).
Finally, a small number of initial conditions
did not converge to any $\FP_\k$ within the simulation time [gray squares in Fig.~\ref{figure3}(b)],
but presumably converged to more exotic states,
e.g., chimeras states consisting of multiple ordered domains~\cite{Panaggio2015},
see SM text for examples.

 \paragraph{Diverging relaxation time.}
We investigated cilia carpets of different size,
and consistently found that the local stability patterns of metachronal waves remain similar to Fig.~\ref{figure2}(b), see SM text.
Similarly, we observe a single dominant wave solution for all system sizes tested, 
with corresponding wave vectors close to $k_\mathrm{I}$ throughout.
Nonetheless, in larger systems, perturbation modes with longer wavelengths and longer relaxation times appear.
The relaxation time $\tau_\mathrm{relax} = \max_i \tau_i$ of the slowest-decaying perturbation mode for the respective dominant wave solution
increases with system length $L=\max(L_x,L_y)$ of the $L_x \times L_y$-simulation domain approximately as
\begin{equation}
\tau_\mathrm{relax} \sim L^2 \quad ,
\label{eq:mermin_wagner}
\end{equation}
see Fig.~\ref{figure2}(c).
While we demonstrate this power law only numerically for cilia carpets,
one can in fact prove this power law analytically for a minimal Kuramoto model with local sinusoidal coupling, see SM text.
This dynamic behavior parallels the Mermin-Wagner theorem from statistical mechanics
for two-dimensional equilibrium systems with continuous symmetries~\cite{Mermin1966}.
For example, in the classical XY model of interacting spins in the plane with short-range interactions,
so-called Goldstone modes appear;
the energy-per-area of these these long-wavelength perturbation modes scales as $1/L^2$ with system length $L$ \cite{Chaikin1995,Mattis1984}.
In a dynamic re-formulation, the relaxation times of these perturbation modes diverge as $\sim L^2$
if we impose over-damped dynamics.
In this sense, one may interpret Eq.~(\ref{eq:mermin_wagner}) as a dynamic Mermin-Wagner theorem of a non-equilibrium system.~\cite{Denes2019a}
The analogy between synchronization and the XY model can be made more explicit
for the classical Kuramoto model with local sinusoidal coupling \cite{Sarkar2021}.

 \paragraph{Synchronization in presence of quenched frequency disorder.}
In real cilia carpets, the intrinsic beat frequencies of individual cilia will slightly differ. 
In a Kuramoto model with all-to-all coupling, 
a second-order phase transition occurs as function of a frequency disorder parameter,
  whereas in Kuramoto models with local coupling a synchronization transition can only be observed in finite systems
\cite{Hong2005, Lee2010}.
 
We now investigate a cilia carpet,
where each cilium has a slightly different intrinsic beat frequency $\omega_i$,
with equation of motion given by Eq.~(\ref{eq:motion}), 
but with $\omega_0$ replaced by $\omega_i$ for cilium $i$, 
i.e., 
$\dot{\varphi}_i = \omega_i - \sum_{j\neq i} \gamma_{ij}\dot{\varphi}_j$. 
Cilia beat frequencies are drawn from a normal distribution with mean $\omega_0$ and standard deviation $\Delta\omega > 0$.
[As a technical point, we rejected frequency sets whose sample standard deviation differed by more than $\approx 1\%$ from $\Delta\omega$.]
We are interested in the synchronization behavior of the cilia carpet as function of $\Delta\omega$, 
averaged over different frequency sets and initial conditions, see SM for details. 

To characterize steady-state solutions, 
we introduce a generalized Kuramoto order parameter, see also~\cite{Gupta2014}
\begin{equation}
\label{eq:rk}
r_\k(\F) = 
N^{-1}
\left|
{\textstyle \sum_j} 
\exp i(\varphi_j + \mathbf{k}\cdot\mathbf{x}_j )
\right|
\quad.
\end{equation}
This order parameter $r_\k$ is close to one,
whenever the cilia phases approximately form a plane traveling wave $\F_k(t)$ with wave vector $\k$,
i.e., $\varphi_j \approx \ol{\varphi} - \k \cdot \x_j$.
The inequality $r_\k(\F) > 2^{-1/2}$
defines mutually disjoint neighborhoods for each $\k$ 
(each of which occupies only a tiny fraction $<10^{-10}$ of the whole phase space).

Fig.~\ref{figure3}(c) shows the fraction of trajectories $\F(t)$ as function of $\Delta\omega$ that both 
(i) converge to the neighborhood of a metachronal wave solution $\FP_k(t)$ with $r_\k[\F(t)] > 2^{-1/2}$, and
(ii) exhibit global \textit{frequency synchronization}, i.e., phase differences between different cilia remain bounded.
This definition for \textit{global metachronal coordination} generalizes a previous definition
for the case $\k=\mathbf{0}$,
which required both `phase cohesiveness' and `frequency synchronization'~\cite{Doerfler2014}.
We find that the fraction of synchronized trajectories 
sharply decreases near a characteristic value of frequency disorder,
$\Delta \omega_{c} / \omega_0 \approx 2.5 \times 10^{-3}$.
This value likely depends on system size, 
as suggested by previous work on two-dimensional Kuramoto models with local coupling ~\cite{Hong2005, Lee2010}.
For intermediate values of $\Delta\omega$ close to the transition point, 
$\Delta\omega \approx \Delta\omega_c$,
we observe a fraction of trajectories that exhibit \textit{partial synchronization}, 
i.e., trajectories satisfy condition (i) [large Kuramoto order parameter],
but not condition (ii) [frequency synchronization], 
apparently because a few cilia did not synchronize and displayed phase drift instead.

 \paragraph{Discussion.}
 We analyzed global stability of metachronal synchronization in cilia carpets
using a multi-scale model,
and found that a single dominant wave solution has a basin-of-attraction
that spans almost the entire phase space of initial conditions
(generalizing early observations for oscillator rings~\cite{Wiley2006}).
The wave direction of this dominant metachronal wave solution encloses an angle of ${\approx}\,60^\circ$
 with the direction of the effective stroke of the cilia beat,
which is close to the experimentally observed value ${\approx}\,90^\circ$,
corresponding to a so-called \textit{dexioplectic} wave~\cite{Machemer1972}.
 The experimentally observed wavelength ${\approx}\,11\,\micron$
is smaller than the wavelength of the dominant wave mode $2\pi/|\k_\mathrm{I}| \approx 34\,\micron$ predicted here;
this discrepancy may simply be a consequence of the cilia density used in our model,
which does not yet allow us to study smaller wavelengths.
 
Linear stability analysis showed that
long-wavelength perturbations of the dominant synchronized state relax only slowly
with relaxation time-scales that increase quadratically with system size.
This dynamic behavior in a non-equilibrium system
parallels the Mermin-Wagner theorem for two-dimensional equilibrium systems with continuous symmetries
(such as the XY models of interacting spins in a plane)~\cite{Mermin1966}.
In these systems, long-wavelength perturbations known as Goldstone modes appear in large systems, 
whose energy-per-area becomes arbitrarily small
and hence their relaxation times diverge if we impose over-damped dynamics.
Noise excites these Goldstone modes, which rules out global order in infinite systems.
Based on the observed divergence of relaxation times, 
we expect a similar behavior for metachronal synchronization in cilia carpets \cite{Solovev2021b}.
 The non-equilibrium dynamics in cilia carpets is thus different
from other non-equilibrium dynamical models such as the Toner-Tu model of flocking birds \cite{Toner1995}:
in that two-dimensional model, global order is possible, 
because the active motion of agents results in a continuous exchange of neighbors.
In contrast, 
the set of neighbors remains fixed in the cilia carpet model.

 Our analysis became possible by a multi-scale simulation approach
that describes beating cilia as phase oscillators~\cite{Friedrich2012,Polotzek2013,Solovev2020a}.
We describe the cilia carpet as an array of phase oscillators,
similar to a Kuramoto model with local coupling~\cite{Dorfler2014},
 yet where direction-dependent coupling functions are calibrated
from detailed hydrodynamic simulations using a measured cilia beat pattern from \textit{Paramecium}
\cite{Machemer1972, Naitoh1984}.
Our approach tries to combine the mathematical elegance of
popular minimal models that idealize beating cilia as orbiting spheres
\cite{Vilfan2006,Guirao2007,Niedermayer2008,Uchida2011,Wollin2011,Friedrich2012,Friedrich2016,Pellicciotta2020,Meng2021}, and
the quantitative predictive power of full-scale numerical simulations that are computationally expensive
\cite{Gueron1999,Elgeti2013,Stein2019}.

 For technical reasons, cilia spacing in our model ($a{=}18\,\micron$)
is larger than in real cilia carpets ($2\,\micron$~\cite{Machemer1972}),
similar to the dilute limit considered in most theoretical studies.
Therefore, we underestimate hydrodynamic interactions,
which are expected to scale as inverse cubed distance of cilia distance in the far field~\cite{Blake1974a,Solovev2020a}.
In dense cilia carpets,
near-field hydrodynamic interactions can change though and even steric repulsion can become important.
  As a consequence, we likely underestimate the characteristic value of disorder of intrinsic beat frequencies
at which synchronization is lost.

Our model could be extended to systems consisting of separated cilia bundles found in airway epithelia~\cite{Pellicciotta2020}.
Future refined models may include internal friction of cilia beating~\cite{Klindt2016,Pellicciotta2020,Nandagiri2020},
and cilia waveform compliance~\cite{Niedermayer2008,Klindt2017},
which are expected to reduce and increase synchronization strength, respectively.
A putative role of basal coupling of cilia contributing to synchronization~\cite{Quaranta2015,Wan2016,Klindt2017}
remains open for cilia carpets,
and has therefore not been included here.
Real cilia carpets are characterized also by
quenched disorder of cilia position,
and non-perfect alignment of cilia~\cite{Guirao2010},
which should reduce the regularity of emergent metachronal waves.
 Intriguingly, some disorder of metachronal coordination might actually be beneficial for transport of suspended particles,
e.g., virus clearance from ciliated airways~\cite{Ramirez2020}.

\begin{acknowledgments}
AS and BMF are supported by the German National Science Foundation (DFG)
through
the \textit{Microswimmers} priority program (DFG grant FR3429/1-1 and FR3429/1-2 to BMF),
a Heisenberg grant (FR3429/4-1),
as well as
through the Excellence Initiative by the German Federal and State Governments
(Clusters of Excellence cfaed EXC-1056 and PoL EXC-2068).
We thank Christa Ringers and Nathalie Jurisch-Yaksi (NTNU), as well as all members of the `Biological Algorithms' group for stimulating discussions.
\end{acknowledgments}

\paragraph{Data availability.}
Python code used to generate results in this manuscript is available in public repositories~\cite{gitall}.
 
\bibliography{cilia_carpet}

\clearpage

\appendix

 \section{Supplemental Material}

\makeatletter
\let\inserttitle\@title
\makeatother

{\noindent
Anton Solovev,
Benjamin M. Friedrich:\\
\textbf{\inserttitle}.\\[0.5mm]
}

\renewcommand{\theequation}{S\arabic{equation}}
\setcounter{equation}{0}  %
\renewcommand{\thefigure}{S\arabic{figure}}
\setcounter{figure}{0}  %
\renewcommand{\thetable}{S\arabic{table}}
\setcounter{table}{0}  %
\renewcommand{\thepage}{S\arabic{page}}
\setcounter{page}{1}  %

\subsection{Numerical methods}

\paragraph{Data availability.}
We deposited code used to generate results in this manuscript as Python packages in three publicly accessible repositories, specifically:
(i) digitalization of three-dimensional cilium beat from stereographic recordings, including coordinate files of the final cilium beat pattern
(ii) routines for generating the triangulated mesh of cilia and boundary surfaces, 
and for solving the hydrodynamic Stokes equation and computing generalized hydrodynamic friction coefficients, (iii) routines for numerical integration of the equation of motion Eq.~(\ref{eq:motion}),
as well as linear stability, global stability, and additional analyses~\cite{gitall}.

\paragraph{Applicability of Stokes equation.}
In the presence of a no-slip boundary surface,
the flow field generated by a static force monopole
decays as $1/d^3$ as function of distance $d$ parallel to the plane in the limit of zero Reynolds number~\cite{Blake1974a}.
For an oscillating force monopole,
whose amplitude oscillates with angular frequency $\omega_0$,
the linearized Navier-Stokes equation predicts that
the 
leading order singularity of
the induced flow field becomes exponentially attenuated beyond a characteristic distance
$\delta = [2\mu/(\rho\omega_0)]^{1/2}$,
where $\mu$ is the dynamic viscosity of the fluid, and $\rho$ its density;
for distances $d\gg\delta$, the flow field decays as $1/d^3$ far from boundaries and as $1/d^5$ close to a plane boundary \cite{Stokes1851,Klindt2015,Wei2019}.
Using a typical cilia beat frequency $\omega_0 / 2\pi = 32\,\mathrm{Hz}$ and parameters for water at room temperature,
we estimate $\delta \approx 100\,\micron$.
Thus, hydrodynamic interactions from nearby cilia should contribute most to synchronization by hydrodynamic interactions.

Additionally,
the flow induced by an oscillating force monopole exhibits a distance-dependent phase lag.
For neighboring cilia, however, this phase lag is small.
Correspondingly, we employ the approximation of zero Reynolds number
and compute the interactions between nearby cilia using the Stokes equation.

\paragraph{Mesh generation.}
Cilia are modeled as slender curved rods with a radius of $0.125\,\micron$ with prescribed centerline,
using a digitalization of cilia beat pattern from unicellular \textit{Paramecium} recorded by~\cite{Machemer1972}
and represented by~\cite{Naitoh1984}.
The simulation geometry representing a local region of a cilia carpet
consists of a boundary surface modeled as a disk of radius $60\,\micron$ represented as a triangular mesh,
whose upper face is coplanar with the $xy$ plane containing the cilia base points $\x_j$.
Triangulated meshes of the shape-changing cilia are anchored to the upper surface of this disk at the respective base points.
For numerical accuracy, we performed local mesh refinement of the  mesh in the vicinity of the base points,
resulting in a mesh with a total of typically $8 \cdot 10^3$ node points,
see Fig~S1(a). This cilia carpet is immersed in an unbounded, Newtonian fluid with dynamic viscosity
$\mu = 10^{-3}\, \textrm{Pa}\,\textrm{s}$  (corresponding to viscosity of water at $20^\circ\,\mathrm{C}$).
For details on mesh generation, see~\cite{Solovev2020a}.

To solve for the surface density of hydrodynamic friction forces
resulting from a shape change of the cilia,
we employ fastBEM, a fast multipole solver for the Stokes equation~\cite{Liu2006}.

\paragraph{Generalized hydrodynamic friction coefficients.}
We compute
hydrodynamic interaction coefficients
$\Gamma_{ij}=\Gamma_{ij}(\varphi_i,\varphi_j)$
in a series of numerical experiments,
where only one cilium with index $j$ beats at a constant frequency $\omega_0$, while other cilia are standing still,
i.e., $\dot{\varphi}_k = 0$ for $k\neq j$.
Using the hydrodynamic solver, we obtain surface force densities $\vec{f}_j(x)$
on the combined surface $\mathcal{S}$ of all cilia and the boundary surface.
We compute the hydrodynamic friction coefficients $\Gamma_{ij}$ as
\begin{equation}
\label{eq:gamma_ij_int}
\Gamma_{ij} =
\int_{\mathcal{S}} \! d^2\x\,
 \frac{\vec{f}_j(\x)}{\omega_0} \cdot
\frac{\partial \x}{\partial\varphi_i}
\comma
\end{equation}
where $\vec{w}_i=\partial{\x}/\partial\varphi_i$ is a rate of displacement of the surface $\mathcal{S}$ corresponding to a change of $\varphi_i$,
while all other $\varphi_k$, $k\neq i$, do not change.
Note that we can restrict the surface integral in Eq.~(\ref{eq:gamma_ij_int}) to the surface $\mathcal{S}_i$ of cilium $i$,
since $\vec{w}_i(\x)=0$ on the rest of the surface $\mathcal{S}\setminus\mathcal{S}_i$.

For each relative orientation of cilia
$\mathbf{d} =\x_j-\x_i$,
we computed
generalized hydrodynamic friction coefficients
$\Gamma_{ij}=\Gamma_{ij}(\varphi_i,\varphi_j)$
characterizing hydrodynamic interaction between cilia.
Specifically, we sampled the respective phases $\varphi_i$ and $\varphi_j$ of the two cilia equidistantly with step size $\Delta\varphi = 2\pi/20$,
while the phases of all other neighboring cilia were set to a constant value of either $0$, $\pi/2$, $\pi$ or $3\pi/2$,
see Fig.~S1(b).  We then averaged over the constant phase the of other cilia,
by fitting a truncated bi-variate Fourier series in $\varphi_i$, $\varphi_j$,
of maximal order $4$
(corresponding to $(2 \cdot 4 + 1) ^2 = 81$ Fourier terms for each $\Gamma_{ij}$).
  In rare cases ($<1\%$), the hydrodynamic solver would unexpectedly fail to converge to the prescribed tolerance ($10^{-3}$);
these data points were excluded from the fit.
The self-friction coefficients $\Gamma_{ii}(\varphi_i)$ are computed in a similar way,
with one cilium phase sampled with step size $\Delta\varphi = 2\pi/20$,
and averaged over a constant phase of its $6$ neighboring cilia
(only $2 \cdot 4 +1 = 9$ terms in Fourier series are kept),
see Fig.~S1(c).
This provided `look-up tables' for subsequent dynamic simulations of the equations of motion of the cilia carpet,
Eq.~\eqref{eq:force_balance}.

While these hydrodynamic simulations consider only a finite cilia array,
they are sufficient to calibrate relevant nearest- and next-to-nearest-neighbor hydrodynamic interactions,
which are later used to simulate larger cilia carpets with periodic boundary conditions.

\paragraph{Visualization of hydrodynamic interaction.}
For Fig.~\ref{figure1}(c), we computed the pairwise normalized hydrodynamic interaction
\begin{equation}
\gamma_{ij}(\vf_i, \vf_j) =   \frac{ \Gamma_{ij}(\vf_i, \vf_j) }{ \Gamma_{ii}(\vf_i) } \quad, 
\end{equation}
using the Fourier sum representation of $\Gamma_{ij}$ and $\Gamma_{ii}$ described above.
For Fig.~\ref{figure1}(d), we computed the root-mean-square average of $\gamma_{ij}(\vf_i, \vf_j)$ as
$\langle \gamma_{ij}^2 \rangle^{1/2}
= (2\pi)^{-1} [ \iint \! d\varphi_i\,d\varphi_j\, \gamma_{ij}^2(\varphi_i,\varphi_j)]^{1/2}$
for nearest and next-to-nearest neighbors.
As a technical point, 
for some next-to-nearest neighbors 
(specifically, for distance $d = \sqrt{3}a$ and direction angles
$\psi = \pm \pi / 6, \,  \pm  5 \pi/6$ relative to $x$-axis,
where $\gamma_{ij}$ is already very small), 
more than $1\%$ but still less than $5\%$ of the hydrodynamic computations did not converge to the prescribed tolerance.
For the visualization of $\langle\gamma_{ij}^2\rangle^{1/2}$ in Fig.~\ref{figure1}(d),
we included all data points in the fit of the Fourier sum, 
including those for which the hydrodynamic computation did not converge.
Note that these problematic next-to-nearest neighbor interactions were not included in the final dynamic computations
because the corresponding hydrodynamic interactions are already very small.

\paragraph{Approximation of pairwise interactions.}
We highlight the two simplifications underlying our effective multi-scale simulation framework.
(i) We introduced a minimal set of effective degrees of freedom, and constrained the full dynamics to these degrees of freedom.
With these constraints imposed, the balance Eq.~\eqref{eq:force_balance} is exact.
(ii) We approximated the $N$-body hydrodynamic interaction as a superposition of pairwise interactions and introduced a distance cut-off.
While the force balance is not exact anymore with these approximations, we numerically confirmed that it still holds to very good accuracy.
Thus, the force balance equation with approximation of pairwise interactions reads
\begin{align}
Q_i(\varphi_i)
& \stackrel{(i)}{=}
\Gamma_{ii}(\varphi_1,\ldots,\varphi_N)\,\dot{\varphi}_i
+ \sum_{j\neq i} \Gamma_{ij}(\varphi_1,\ldots,\varphi_N)\,\dot{\varphi}_j
\notag \\
& \stackrel{(ii)}{\approx}
\Gamma_{ii}(\varphi_i)\,\dot{\varphi}_i
+ \sum_{j\in \N_i} \Gamma_{ij}(\varphi_i,\varphi_j)\,\dot{\varphi}_j
\quad.
\end{align}
Here, $\N_i$
is the set of neighbors of cilium~$i$,
which includes all six nearest neighbors (at distance $a=18\,\micron$)
and two next-to-nearest neighbors located at $\pm d\,\e_y$ with $d=\sqrt{3}a$
(corresponding to direction angle $\psi=\pm\pi/2$),
i.e., located along direction of the cilia effective stroke,
where hydrodynamic interactions are the strongest, see Fig.~\ref{figure1}(d). Next-to-nearest neighbor interactions along the other directions are much weaker,
and were therefore not included in the final simulations for reasons of computational performance.
Initial simulations showed that including these interactions with next-to-nearest neighbors
only slightly changed quantitative results, and did not affect any of our qualitative conclusions.

\begin{figure}
\includegraphics[width=\linewidth]{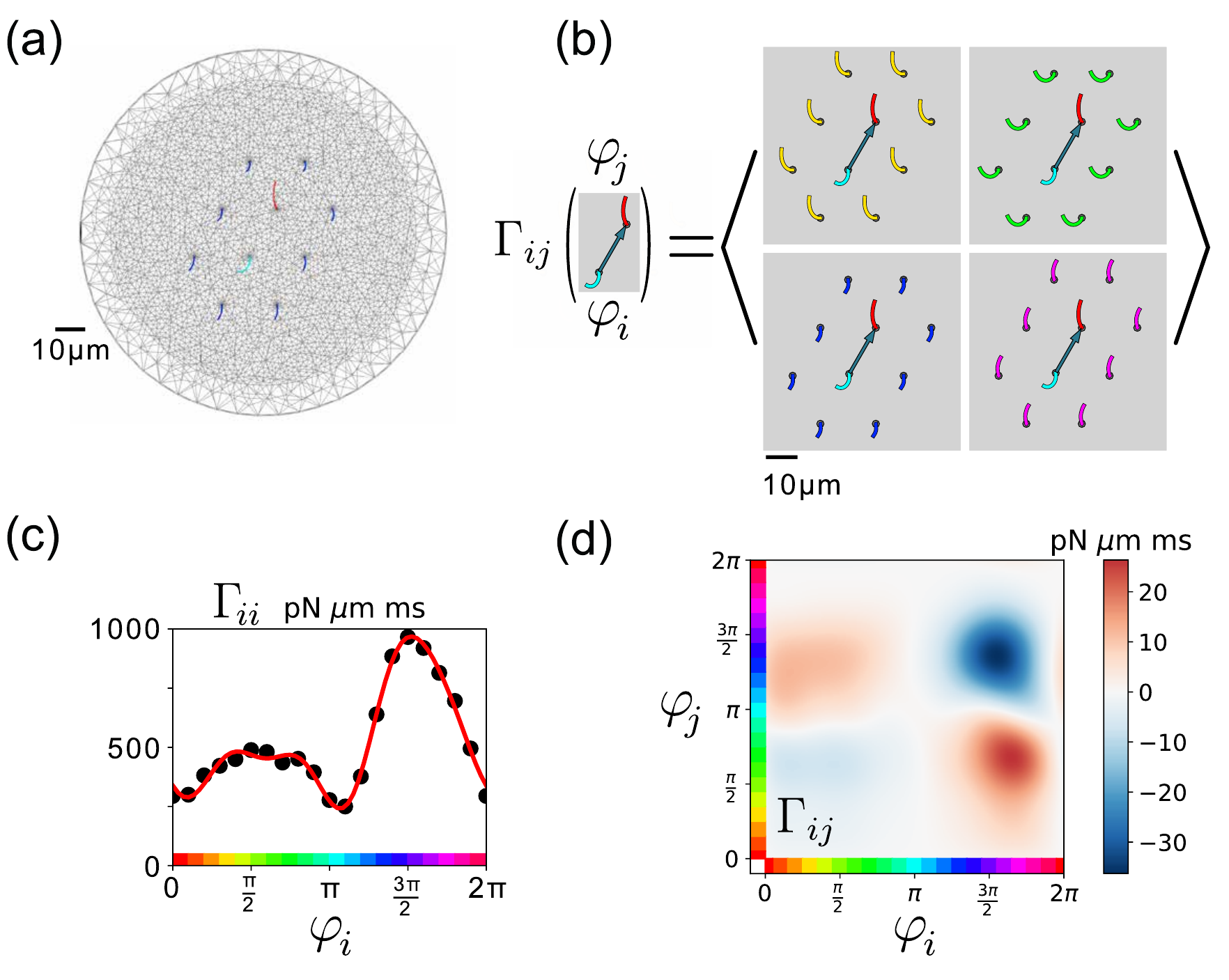}
\caption{
\textbf{Computation of generalized hydrodynamic friction coefficients.}
(a) Top view on the triangulated mesh representing cilia and boundary surface as used in hydrodynamic computations.
(b) Illustration of the method used to average out the phases of those surrounding cilia that are not directly involved in the interaction pair $(i,j)$:
we obtain $\Gamma_{ij}$ as a function of only $\varphi_i$ and $\varphi_j$ by averaging over a constant value
of $\varphi_k$ for $k\neq i,j$.
(c)
Self-friction coefficient $\Gamma_{ii}$ as function of cilium phase $\varphi_i$.
Dots represent values of $\Gamma_{ii}$ directly obtained from hydrodynamic computations.
    The solid line represents the fitted Fourier series used as `look-up table' in all subsequent dynamic computations.
(d)
Hydrodynamic interaction $\Gamma_{ij}$ 
as function of cilia phases $\varphi_i$ and $\varphi_j$ (Fourier sum fit).
Separation vector of cilia bases, $\x_j-\x_i=a\,(\cos\psi\,\e_x + \sin\psi\,\e_y)$, $\psi{=}\pi/3$
(same as in main text).
}
\label{figureS1}
\end{figure}

\paragraph{Active cilia driving forces.}
For our choice of reference condition,
the active driving forces $Q_i(\varphi_i)$ are given by
\begin{equation}
Q_i(\varphi_i) = \omega_0 \, \Gamma_{ii}(\varphi_i) \quad,
\end{equation}
corresponding to a single cilium
that beats at a constant frequency
(while its neighbors are at rest and only act as obstacles for the fluid).
\paragraph{Equation of motion.}
Numerically, we solve the equation of motion Eq.~(\ref{eq:motion}) in the form
\begin{equation}
\label{eq:eq_motion}
 \dot{\F} = \Gmat^{-1} \cdot \Q \quad.
\end{equation}
The coupling functions $\G_{ij}$ depend only on the phases $\varphi_i$ and $\varphi_j$ and the relative positions of cilia $i$ and $j$,
allowing for efficient storage.

Alternatively, we could introduce the generalized mobility matrix $\Mmat = \Gmat^{-1}$, and
the vector of active driving forces $\Q$ with components $Q_j(\varphi_j)$.
The equation of motion
$\dot{\F} = \Mmat \cdot \Q$
can then be written as a system of $N$ coupled phase oscillators
\begin{equation}
\dot{\varphi}_i = \omega_0 + \sum_{j \neq i} c_{ij}(\varphi_1,\ldots,\varphi_N) \quad,
\end{equation}
with coupling functions
$c_{ij} = (\Mmat\cdot\Q)_{ij} - \omega_0\,\delta_{ij}$.
Diagonal entries $c_{ii}$ characterize a modulation of beat frequency due to the presence of nearby cilia.
As consequence of the no-slip boundary surface,
hydrodynamic interactions decay with inverse cubed distance close to the surface~\cite{Blake1974a}.
Thus, in the limit of low cilia density with $\ell \ll a$
where $\ell$ denotes cilia length,
we have $c_{ij} \sim (\ell /a)^3$
for neighbor cilia with $j\in \N_i$.
Yet, even for $j\notin \N_i$,
$c_{ij}$ is in general non-zero albeit small, decaying at least as $(\ell /a)^6$.
Thus, although the generalized friction matrix $\Gmat$ is sparse
(given the approximation of including only nearest-neighbor interactions),
the generalized mobility matrix $\Mmat$ will be non-sparse in general.

 Eq.~\eqref{eq:eq_motion} represents a generalized Kuramoto model with local coupling.
Indeed, if we set $c_{ij} = \varepsilon\,\sin(\varphi_i-\varphi_j)$ for nearest neighbors, and $c_{ij}=0$ else, 
we would obtain the classical Kuramoto model with local sinusoidal coupling in two space dimensions.
 
\paragraph{Numeric integration of equation of motion.}
We used a 4(5)-Runge-Kutta scheme with adaptive time-step (Python package \textit{scipy})
to numerically integrate the deterministic equation of motion, Eq.~\eqref{eq:eq_motion}.
We used numerical tolerance $10^{-8}$ to determine fixed points and Lyapunov exponents from the linear stability analysis,
and a numerical tolerance of $10^{-6}$ for all other computations.
Intersections with the Poincar\'e plane $H$ defined by $\ol{\varphi}=0$ were detected using the integrated event handler.
 In each time-step, we compute the right side of the equation of motion $\dot{\F}=\Gmat^{-1}\cdot \Q$  using a sparse linear solver.

\begin{figure}
\includegraphics[width=\linewidth]{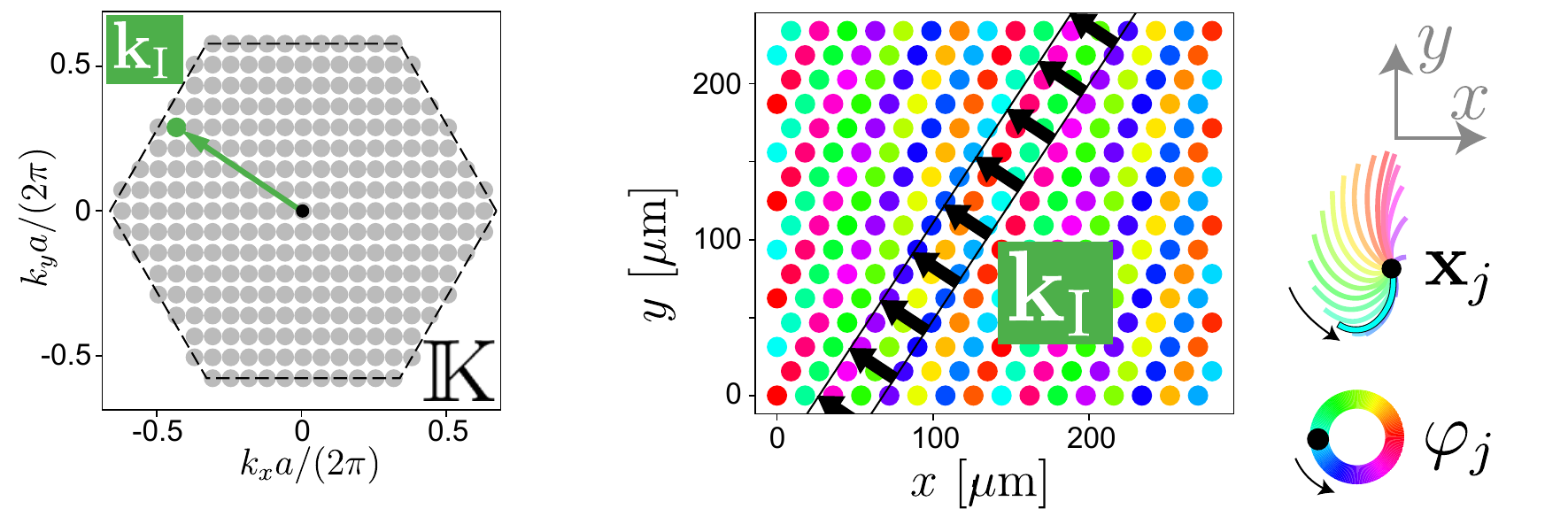}
\caption{
\textbf{Visualization of dominant wave mode $\k_\mathrm{I}$.}
\textit{Left:}
Position of wave mode $\k_\mathrm{I}$ in the Brillouin zone of admissible wave vectors
for the case of a $16\times 16$ cilia carpet.
\textit{Right:}
Corresponding traveling wave:
colored dots at triangular lattice positions of cilia base points $\x_j$
represent respective cilia phase $\varphi_j = - \k_{\mathrm{I}} \cdot \x_j$ according to the color wheel;
$\k_\mathrm{I} \, a/ (2 \pi) = (- 7 / 16, \sqrt{3} / 6)$.  }
\label{figureS2}
\end{figure}

\paragraph{Reciprocal lattice of metachronal wave vectors and Brillouin zone.}
We introduce basis vectors
$\vec{d}_x$ and $\vec{d}_y$ of the reciprocal lattice
defined by a tiling of the plane by copies of the unit cell of $N$ cilia
\begin{equation}
\vec{d}_x =  \left(\begin{array}{c} \frac{2\pi}{L_x} \\ 0\end{array} \right)
 , \quad
 \vec{d}_y =
 \left(\begin{array}{c} 0 \\ \frac{2\pi}{L_y} \end{array} \right)
  \quad,
\end{equation}
where $L_x = N_x\, a$ and $L_y = \sqrt{3} N_y\, a /2 $ denote the length of the rectangular unit cell in $x$ and $y$ direction, respectively.
Any wave vector $\k$ in the reciprocal lattice can be written as
\begin{equation}
\vec{k} = n_x \vec{d}_x + n_y \vec{d}_y = k_x\vec{e}_x + k_y \vec{e}_y
\quad,
\label{eqn:k_allowed}
\end{equation}
with integers
$n_x, n_y \in\mathbbm{Z}$,
or, alternatively,
with vector components
$k_x = n_x\,2\pi / L_x$
and
$k_y = n_y\,2\pi / L_y$
with respect to the normalized unit vectors $\e_x = (1,0)^T$ and $\e_y=(0,1)^T$.
The regular spacing of cilia at lattice positions $\x_j$ inside the unit cell defines a Brillouin zone $\mathbbm{K}$:
in the case of a triangular lattice, this Brillouin zone can be chosen as a hexagon with edge length $k_{\max} = 4\pi / (3a)$,
see  Fig.~\ref{figure2}(a).
This Brillouin zone contains $N=|\mathbbm{K}|$ unique wave vectors.
Any other wave vector $\k'$ of the reciprocal lattice
can be mapped either inside or on the border of this hexagon
using the equivalence relation
$\exp (i\,\k' \cdot \x_j) = \exp(i\,\k \cdot \x_j)$ for all $j$.
A visualization of the dominant wave mode $\k_\mathrm{I}$ is shown in Fig.~\ref{figureS2}.

For a classical Kuramoto model with sinusoidal nearest-neighbor coupling,
each wave vector $\k \in \K$ defines a periodic solution $\F_\k$
with components $\varphi_i = \omega_0 t - \mathbf{k}\cdot\mathbf{x}_i$
(also called $k$-twist~\cite{Peruani2010,Wetzel2017} or \textit{splay states}~\cite{Strogatz1993} in one-dimensional oscillator chains),
 see also section on the Kuramoto model below.
 For the cilia carpet model considered in the main text,
we find periodic solutions that deviate slightly from these perfect traveling waves.

\paragraph{Numeric search for periodic solutions.}

To find periodic solutions $\F_\k^\ast(t)$
of the generalized Kuramoto model given by Eq.~\eqref{eq:force_balance},
we numerically searched in the vicinity of the periodic solutions $\F_\k(t)$ of the classical Kuramoto model.
Specifically, we searched for fixed points $\FP$ of the Poincar{\'e} map $\Lmap$
for the Poincar{\'e} plane $H$ given by $\ol{\varphi}=0$,
where $\ol{\varphi}=\sum_j \varphi_j/N$ denotes the global phase
\begin{align}
\Lmap \,:\,
& H \rightarrow H \notag \\
& \F_0 \longmapsto  \F_1 -  2 \pi\,\vec{1}
\quad.
\end{align}
Here,
$\F_0 = \F(t_0)\in H$
is the start point of a trajectory $\F(t)$
that intersects the shifted Poincar{\'e} plane $H+2\pi\,\vec{1}$ at
$\F_1 = \F(t_1)$, i.e., $\gp(t_0)=0$ and $\gp(t_1)=2\pi$.
Numerically, it turned out to be easier to start also with initial phase vectors
that had a non-zero global phase, i.e.,
$\gp(t_0)=\varphi_0$ and $\gp(t_1)=2\pi + \varphi_0$.
We found fixed points $\FP$ by numerically searching for zeros of the following vector function,
where the last term effectively restricts the search to the Poincar{\'e} plane $H$
\begin{equation}
\mathbf{D}(\F_0) = \Lmap(\F_0) - \F_0  - \gp (\F_0) \vec{1}\quad.
\end{equation}
Note that the condition $\mathbf{D}(\F_0)=\zero$ actually implies
both $\Lmap(\F_0) - \F_0=\zero$ \textit{and} $\gp (\F_0)=0$.
Hence,
$D(\FP) = \vec{0}$ yields a fixed point $\FP\in H$ with zero global phase.
By running the numerical search algorithm $N$ times
with start vectors $\F_0$ given by plane waves $\varphi_{i} = - \k\cdot\x_i$ for each $k\in\mathbbm{K}$,
we found $N$ different fixed points $\FP_\k$.
The Kuramoto order parameters $r_\k$ defined in Eq.~\eqref{eq:rk}
evaluated at the fixed points almost equal one with $ r_\k(\FP_k) > 1 -  2 \cdot 10^{-3}$.
This confirms that these fixed points correspond to periodic solutions $\FP_\k(t)$ that are indeed close to perfect traveling waves.

\paragraph{Linear stability analysis}
We numerically find the linearized Poincar{\'e} map $\Lmat_k$ near a fixed point $\FP_\k$ [see Eq.~\eqref{eq:Lmat}],
by computing the Poincar{\'e} map $\L$ for small perturbations.
Specifically, we apply small perturbations $\D_{0}^{(i)}$
with $| \D_{0}^{(i)} | = 10^{-2} N^{1/2}$ and zero global phase
in $N-1$ linearly-independent directions,
hence $\FP_\k + \D_{0}^{(i)} \in H$.
We then compute
\begin{equation}
\D_{1}^{(i)} = \Lmap(\FP + \D_{0}^{(i)}) - \FP, \quad i=1\ldots N-1 \quad.
\end{equation}
In order to obtain a $N$-dimensional matrix representation of $\Lmat$,
this $N-1$-dimensional set of perturbations is complemented by normal vector to the Poincar{\'e} plane,
$\D_{0}^{(N)} = \D_{1}^{(N)} = \vec{1}\approx \dot{\F}^{*}_\k / | \dot{\F}^{*}_\k |$.

\begin{figure}
\includegraphics[width=\linewidth]{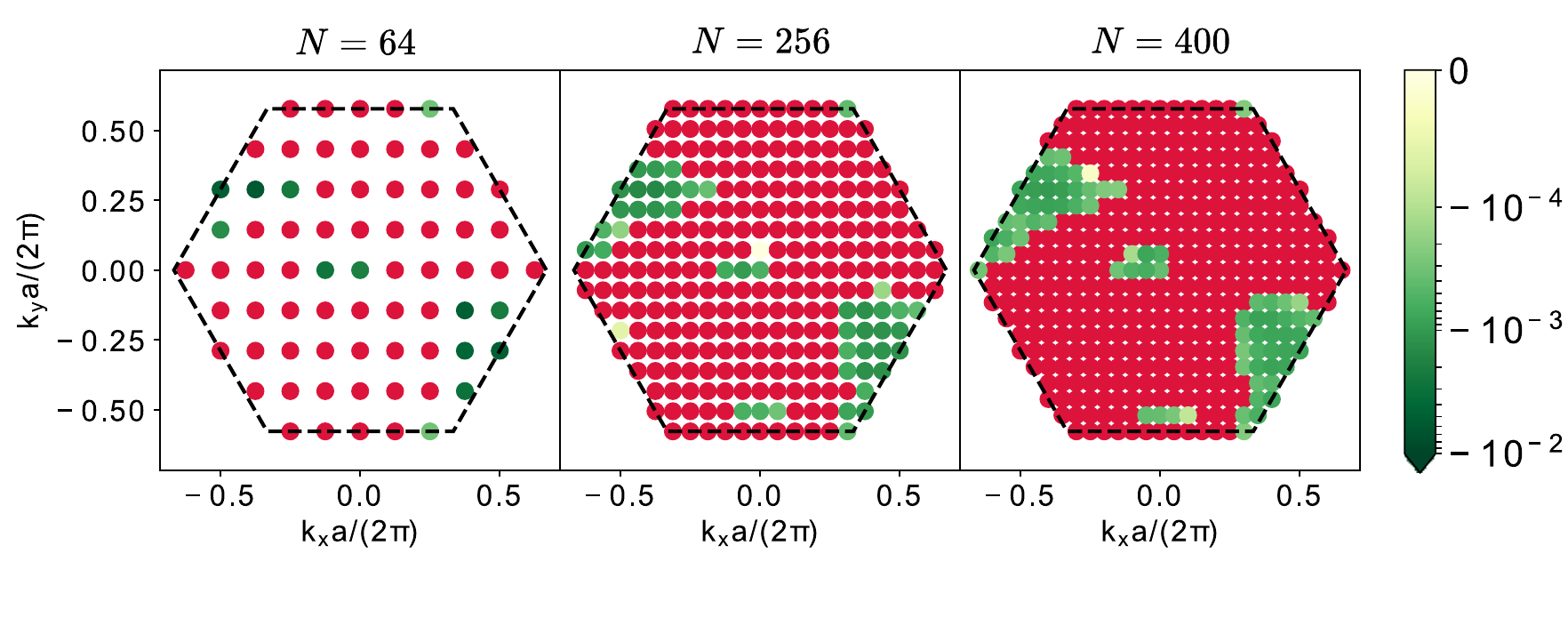}
\caption{
\textbf{Linear stability analysis for systems of different size.}
We performed linear stability analyses for each wave vector $\k$ inside a Brillouin zone for systems of different sizes
similar to Fig.~\ref{figure2}(b) in the main text.
In all three cases, stability patterns are similar:
\textit{left:} $8 \times 8$ carpet with $N=64$ cilia,
\textit{middle:} $16 \times 16$ carpet with $N=256$ cilia,
\textit{right:} $20 \times 20$ carpet with $N=400$ cilia.
Green colors represent $\mathrm{max}\,\Re\lambda_j$ of respective Lyapunov exponents $\lambda_j$ for linearly stable wave modes $\k$;
red dots represent modes that are linearly unstable.
The absolute values of eigenvalues tend to zero as system size increases, 
as discussed in Fig.~\ref{figure2}(c) in the main text
(which reports the relaxation time
$\tau_\mathrm{relax}= 2\pi / |\omega_\k \max\,\Re\lambda_j|$).
}
\label{figureS3}
\end{figure}

By Eq.~\eqref{eq:Lmat}, we expect
$
\D_{1}^{(i)} = \Lmat \cdot \D_{0}^{(i)}
$
for $i=1,\ldots,N$.
We introduce matrices
$\mathbf{D}_0$ and $\mathbf{D}_1$
that comprise the
$N$ perturbation column-vectors $\D_{0}^{(i)}$, and
the $N$ response column-vectors $\D_{1}^{(i)}$, respectively,
as
$\mathbf{D}_k = \left(\D_{k}^{(1)}, \D_{k}^{(2)},\ldots,\D_{k}^{(N)} \right)$ for $k \in\{0,1\}$.
Thus, $\mathbf{D}_1 = \Lmat \cdot \mathbf{D}_0$
and the linearized Poincar{\'e} map matrix is found as
$
\Lmat = \mathbf{D}_1 \cdot \mathbf{D}_{0}^{-1}
$.
Fig.~\ref{figureS3} shows results of a linear stability analysis for cilia carpets of different sizes.

\paragraph{Basins-of-attraction.}
To estimate the relative size of basins-of-attractions,
we computed $n=400$ trajectories with initial conditions given by uniformly sampled random phase vectors.
For each trajectory, we integrated the equation of motion Eq.~(\ref{eq:motion}) for $m=4000$ beat cycles
(corresponding to an integration time $t_m \approx 4000\,T_0$).
All $n$ trajectories converged to a neighborhood of a wave $\F_\k$ for a suitable wave vector $\k$, 
as determined by a Kuramoto order parameter $r_{\k} > 0.99$.
Additionally, we observed that each of the $n$ trajectories apparently converged to a fixed point, 
by checking that the Euclidean norm of the change of the phase vector 
$d_l=N^{-1/2} \left\Vert\,\L^l[\F(0)] - \L^{l-1}[\F(0)]\,\right\Vert_2$ 
during one beat cycle was decreasing and
sufficiently small
after $m=4000$ cycles, $d_m < 2 \cdot 10^{-4}$.

In fact,
all trajectories converged to just five waves
(all of which are very close to each other in terms of both wave direction and wavelength);
the majority of trajectories converged to either $\k_\mathrm{I}$ ($86\%\pm2\%$) or $\k_\mathrm{II}$ ($13\%\pm2\%$)
(introduced in the main text Fig.~\ref{figure2}(a)).
The error $e_\k$ was computed as the standard error of a Bernoulli trial~\cite{Menck2013}
\begin{equation}
e_\k = \sqrt{\frac{B_\k (1 - B_\k)}{n}},
\end{equation}
where $B_{\k} \in [0,1]$ is the relative size of the basin-of-attraction, and $n$ is the total number of trajectories.

\paragraph{Slice-visualization of basins-of-attraction.}
In Fig.~\ref{figure2}(b), we additionally visualize convergence
for a specific set of initial conditions of the form
$\varphi_j = - \vec{m}  \cdot \x_j$,
with ``off-lattice'' wave vectors $\vec{m} \not\in \K$
(thus  $\vec{m}$ does not necessarily respect the periodicity of the lattice).
Note that these special trajectories were not used in calculating the relative size of the basins-of-attraction,
as the initial conditions were not drawn randomly.

\begin{figure}
\includegraphics[width=\linewidth]{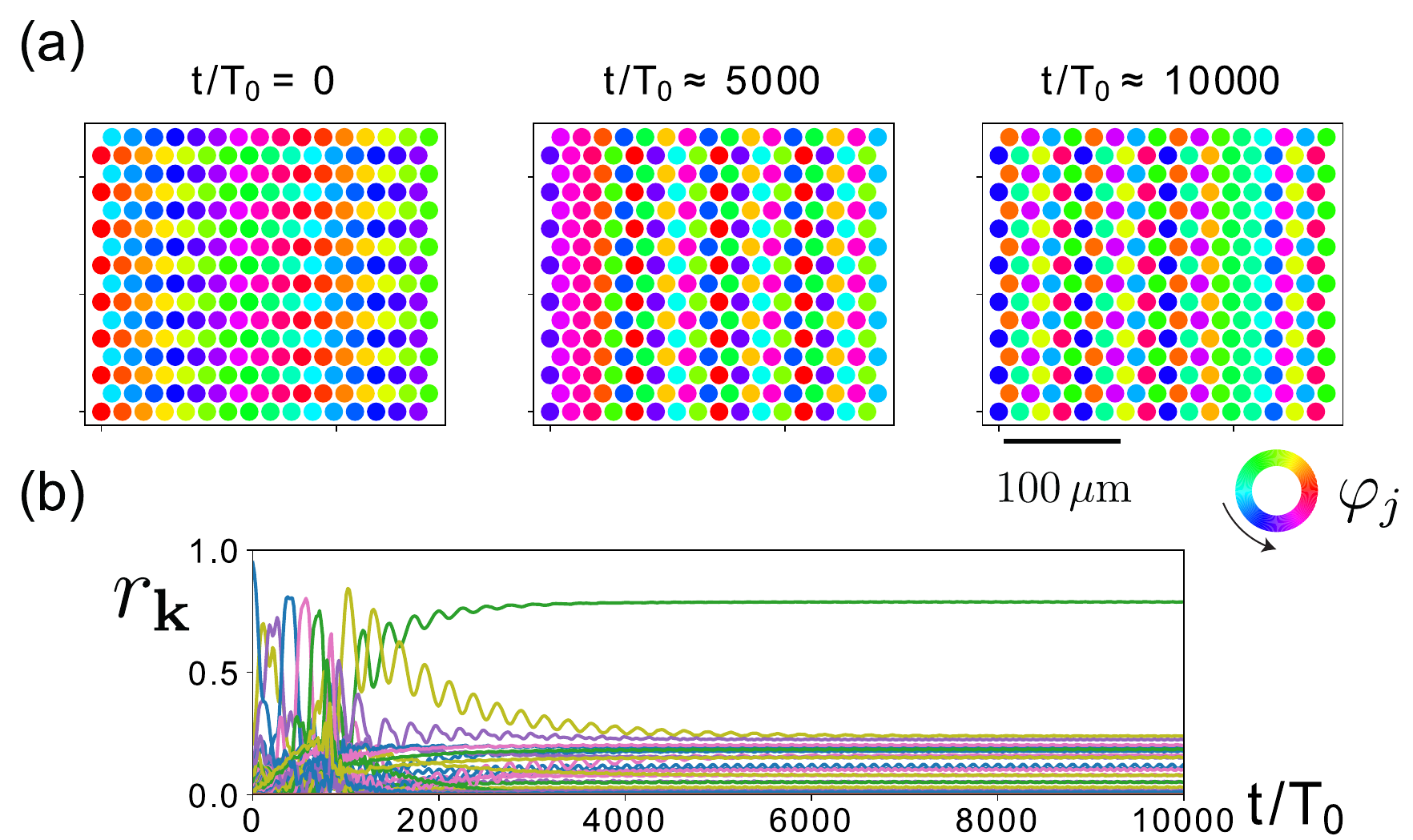}
\caption{
\textbf{Chimera states for special initial conditions.}
(a)
Visualization of phase vectors for noise-free dynamics at different times:
Colored dots represent cilium phase at respective lattice position according to the color wheel.
\textit{Left:}
Initial condition:
off-lattice wave $\varphi_j(t=0) = - \l \cdot \x_j$ with $\l \, a / ( 2\pi) = (- 2 / 39, - \sqrt{3} / 3)$.
\textit{Middle, right:}
At times $t\approx 5 \cdot 10^{3}\, T_0$ and $t\approx 10^{4}\, T_0$,
we observe co-existence of two ordered sub-domains, resembling a chimera state.
(b)
For the same trajectory,
we plot order parameters $r_\k$ for every $\k \in \K$.
For $t \approx 5 \cdot 10^3 \, T_0$, the order parameters reached steady state.
}
\label{figureS4}
\end{figure}

Each of the trajectories was integrated
until it converged to one of the fixed points $\FP_{\k}$
(using the same numerical convergence criterion as detailed in the section \textit{Basins-of-attraction} above).
Intriguingly, for some of the initial conditions, 
trajectories did not converge do any of the fixed points $\FP_{\k}$
[gray squares in Fig.~\ref{figure2}(c)].
Some of these special trajectories apparently became attracted to some other fixed point $\FP$
different from any of the $\FP_\k$, $\k\in\K$
(i.e., the distances $d_l$ introduced in the section above approached zero,
but all Kuramoto order parameters remained below the threshold at the end of the integration time, $r_\k<0.99$ for all $\k$).
Other special trajectories were not attracted to any fixed point $\FP$ even after a long integration time $t \approx 10^4 \,T_0$.
Nonetheless, the dynamics of these later trajectories had become stationary in the sense that
the Kuramoto order parameters $r_\k$, $\k \in \K$, either did not change in time anymore or oscillated in a regular way.
Visual inspection revealed that
these initial conditions had become attracted to more exotic states,
such as chimera states (i.e., states with at least two ordered sub-domains)~\cite{Panaggio2015},
see Fig.~\ref{figureS4} for an example. However, the combined relative size of the basins-of-attraction of these exotic states is negligible;
therefore, these states are not in focus of this study.

As expected, the basin of the dominant wave vector $\k_{\mathrm{I}}$ comprises a large portion of initial conditions 
in the slice of phase space shown in Fig.~\ref{figure3}(b).
Specifically, most initial conditions corresponding to unstable waves vectors $\k$ became attracted to the dominant wave $\k_\mathrm{I}$.
On the other hand, initial conditions in the vicinity of a stable wave vector $\k$ different from $\k_\mathrm{I}$
are likely to become attracted to this wave $\k$, see the magnified region in Fig.~\ref{figure3}(b).

\section{Quenched frequency disorder}
In Fig.~\ref{figure3}(c), we show the fraction of synchronized trajectories as a function of 
a frequency disorder parameter $\Delta \omega$.
Specifically, we drew sets of random intrinsic beat frequencies
$\Omega = (\omega_1, \ldots, \omega_N)$,
where the intrinsic frequency $\omega_i$ of cilium $i$ 
was drawn from a normal distribution with mean $\omega_0$ and standard deviation $\Delta\omega$.
As a technical point, the (biased) sample variance
\begin{equation}
\text{Var}(\Omega) = \frac{1}{N} \sum_i \left[ \omega_i-\mu(\Omega) \right]^2\quad,
\end{equation}
where
$\mu(\Omega) = N^{-1} \sum_i \omega_i$
denotes the sample mean, 
may vary from its expectation value $\Delta \omega^2$.
We rejected frequency sets, where $\text{Var}(\Omega)^{1/2}$ differed from $\Delta\omega$ by more than $1\%$.
Without this rejection (which amounts to about $82\%$ of frequency sets),
the synchronization transition in Fig.~\ref{figure3}(c) would appear more gradual.

For each value of $\Delta \omega > 0 $ considered,
we first generated $m=25$ valid frequency sets. For each frequency set, we then integrated $n=10$ trajectories, starting from a fixed sub-sample of initial conditions.
This sub-sample had been selected before from a larger sample of random initial conditions (uniformly distributed),
such that the previously determined relative sizes of basins-of-attraction for the case without frequency disorder was faithfully reproduced
(for nine initial conditions, the trajectories converged to wave $\ki$ and for one initial condition, the trajectory converged to $\kii$ for $\Delta\omega=0$).
Using only a small number of initial conditions reduced computation times considerably.

For different frequency set $\Omega$,
periodic solutions of the system (and corresponding fixed points of the Poincar{\'e} map)
will slightly differ from the periodic solutions found for the case $\Delta \omega = 0$.
Therefore, 
in order to compute the relative size of basins-of-attractions in Fig.~\ref{figure3}(c),
we employ a sufficiently large neighborhood of the plane wave solution $\F_\k(t)$.
More precisely, we say that a trajectory $\F(t)$ \textit{synchronized to wave $\k$} if
the following two conditions are satisfied
\begin{itemize}
\item[(i)]
The respective Kuramoto order parameter was large at the end of the integration time, $r_\k > \sqrt{2}/2$.
\item[(ii)]
$\F(t)$ converges to a fixed point of the Poincar{\'e} map.
\end{itemize}
Each of the trajectories was integrated
until condition (ii) was met or a maximum integration time $t \approx 1.6 \times 10^{4}\,T_0$ was reached.
To check convergence to a fixed point,
the same criterion as in section `\textit{Basins-of-attraction}' was used.

Without frequency disorder, $\Delta \omega = 0$,
conditions (i) and (ii) are essentially equivalent,
except for few rare cases, where initial conditions converged to exotic states, e.g., chimera states.
However, in the case of frequency disorder with $\Delta \omega > 0$,
the two conditions (i) and (ii) are no longer approximately equivalent,
and we observe trajectories that satisfy condition (i) but not (ii),
especially close to the synchronization transition.
We refer to these trajectories with \textit{partial synchronization}
[red color in Fig.~\ref{figure3}(c)].

\section{Kuramoto models with local coupling}

For the convenience of the reader, we review basic facts on the classical Kuramoto model with local coupling,
part of which can be found in the standard literature~\cite{Pikovsky2003}.

\subsection{One-dimensional chain of phase oscillators with nearest-neighbor sinusoidal coupling}

We consider a one-dimensional chain of
$N$ coupled phase oscillators with periodic boundary conditions.
The oscillators in this ring topology are supposed to have equal angular frequency $\omega_0$
and are coupled to their neighbors by a symmetric sinusoidal coupling
with total coupling strength $K$
\begin{align}
\dot{\varphi}_j
= \omega_0
  & + \frac{K}{2}\, \sin(\varphi_{j-1} - \varphi_j ) \notag \\
  & + \frac{K}{2}\, \sin(\varphi_{j+1} - \varphi_j ), \quad j=1,\ldots, N \quad.
\label{eqn:dphidt-sine}
\end{align}
For notational convenience, oscillator indices are considered modulo $N$
(i.e., oscillator number $N$ is coupled again to oscillator number $1$).
We assume a positive synchronization strength $K>0$;
correspondingly, the in-phase synchronized state is stable.

Traveling waves with angular wave number $k$
define periodic solutions   $\FP_{k} = \F_{k}$
\begin{equation}
\F_k : \varphi_j(t) = \omega_0 t - k j , \quad j=1,\ldots,N\quad,
\label{eqn:phij_plane_wave}
\end{equation}
where $k=2\pi\,m/N$ for some integer $m\in\mathbbm{Z}$.

The fundamental perturbation modes of the Poincar{\'e} map for these periodic solutions $\F_k$
are simply the Fourier modes for the chain
with angular wave number $\nu$
\begin{equation}
\label{eq:perturbation}
\boldsymbol{\Delta}_\nu : \delta_j = \exp\left( - i \,\nu j \right) \quad,
\end{equation}
where $\nu=2\pi\,n/N$ for some $n=1,\ldots,N-1$ ($n=0$ would correspond to a trivial phase shift).
The corresponding eigenvalues of the linearized Poincar{\'e} map $\ln L$,
which we call dimensionless Lyapunov exponents,
read
\begin{equation}
\lambda_{k\nu} = - K T_0  \left( 1 - \cos{\nu}\right) \cos{k}
\quad.
\label{eq:lambda_ring}
\end{equation}
This can be proven by substituting the perturbation Eq.~(\ref{eq:perturbation})
and keeping only terms to linear order. 
The periodic solution for wave number $k$ is linearly stable if and only if the real parts of all eigenvalues $\lambda_{k\nu}$ are strictly negative;
hence, according to Eq.~\eqref{eq:lambda_ring},
exactly the solutions with
$|m| < \lfloor N /2 \rfloor$ are linearly stable.

We can now read off the dimensionless Lyapunov exponents of the slowest decaying mode for each \textit{stable} periodic solution
and find
\begin{align}
\max_{\nu\neq 0} \lambda_{k\nu}
& = -  K T_0 \left(1 - \cos{\frac{2 \pi }{N}}\right) \cos{k}
\notag \\
& \approx - K T_0 \frac{4 \pi^2}{N^2} \cos{k} \sim N^{-2} \sim L^{-2}
\quad.
\end{align}
Here, we introduced a system length $L = N a$, where $a$ is the spacing between oscillators.
Thus, the long wavelength perturbations ($|\k| \to 0$) are indeed those that decay the slowest,
with a decay rate that scales as the inverse square of system length $L=N a$.

In the main text,
we describe a similar scaling
for the relaxation time $\tau_\mathrm{relax}$,
which is inversely proportional to Lyapunov exponent $\max\,\lambda_j$ of the slowest decaying perturbation mode,
for the periodic solution $\FP_{\k_\mathrm{I}}(t)$ corresponding to the dominant wave mode $\k_\mathrm{I}$,
see Fig.~\ref{figure2}(d).
In addition, we numerically checked that the largest dimension $L=\max{(L_x,L_y)}$
dominates the scaling also if $N_x \neq N_y$
(both for the Kuramoto and the cilia carpet models).

\subsection{Dispersion relation for the one-dimensional Kuramoto model with local coupling}
\newcommand*{\fr}[2]{\frac{2 \pi #1} {N} #2 }
\newcommand*{\co}[2]{\cos \fr{#1}{#2} }

As a generalization of Eq.~(\ref{eqn:dphidt-sine}), we can consider 
the Sakaguchi-Kuramoto model with local coupling~\cite{Sakaguchi1986}
$c_{i j } (\varphi_i, \varphi_j) = \varepsilon\sin(\varphi_j-\varphi_i+\delta)$, 
i.e, with additional phase shift $\delta$ (as introduced in the main text).
This generalized one-dimensional Kuramoto model can be written as
\begin{align}
    \label{eq:sine_cosine_eq_motion}
    \dot{\varphi}_j
    = \omega_0
    & + \frac{K}{2}\, \sin(\varphi_{j-1} - \varphi_j)
	& + \frac{U}{2}\, \cos(\varphi_{j-1} - \varphi_j) \notag \\
	& + \frac{K}{2}\, \sin(\varphi_{j+1} - \varphi_j) 
    & + \frac{U}{2}\, \cos(\varphi_{j+1} - \varphi_j),
    \notag \\ & \quad j=1,\ldots, N \quad,
\end{align}
where $K=2\varepsilon\cos\delta$ and $U=2\varepsilon\sin\delta$.
We make an Ansatz of traveling waves
\begin{equation}
    \label{eq:sine_cosine_wave}
    \F_k : \varphi_j(t) = \omega_k t - k j , \quad j=1,\ldots,N \ ,
\end{equation}
with frequencies $\omega_k$ and $k=2\pi\,m/N$ for some integer $m\in\mathbbm{Z}$.
Substituting this Ansatz into Eq.~\eqref{eq:sine_cosine_eq_motion},
yields periodic wave solutions with frequencies $\omega_k$ with
\begin{equation}
\label{eq:dispersion_kuramoto}
\omega_k / \omega_{k=0} = 1 + \beta ( \cos k-1) \quad,
\end{equation}
where $\omega_{k=0} = \omega_0 + U$ and $\beta=U/\omega_{k=0}$, 
Eq.~\eqref{eq:dispersion_kuramoto}.
Thus, an additional cosine term in the coupling function $c_{ij}$ causes a characteristic frequency dispersion relation. 
The stability of wave solutions, however, is not altered, as can be shown analogous to the previous section. 

\section{Kuramoto model with nearest-neighbor sinusoidal coupling in $d$ dimensions}

More generally,
we can consider a Kuramoto model of phase oscillators with identical frequencies on a cubic lattice
with lattice spacing $a$ and lattice positions $\x_i$ in $d$-dimensional space
and local sinusoidal coupling.
Each oscillator with phase variable $\varphi_i$ is coupled to its $2d$ nearest neighbors (enumerated by an index set $\N_i$)
with total coupling strength $K$
\begin{equation}
\label{eq:Kuramoto_sine}
\dot{\varphi}_i = \omega_0 - \frac{K}{2d} \sum_{j\in\N_i} \sin(\varphi_i-\varphi_j) \quad.
\end{equation}
We assume periodic boundary conditions with system size $N_1 \times \ldots \times N_d$.

Linear stability analysis yields a set of fundamental perturbation modes
\begin{equation}
\label{eq:delta_l}
\boldsymbol{\Delta}_\l : \delta_j = \exp( - i\, \l\cdot\x_j) \text{ for } \l\in\mathbbm{K}\setminus\{\zero\}
\end{equation}
with corresponding dimensionless Lyapunov exponents
\footnote{
For the calculation, note
\begin{align}
\dot{\varphi}_i
&= \omega_0 - \frac{K}{2d} \sum_{j\in\N_i} \sin( \varphi_{\k,i}+\varepsilon\,\Delta_{\l,i} - \varphi_{\k,j}-\varepsilon\,\Delta_{\l,j} ) \notag \\
&= \omega_0 -
  \frac{K}{2d} \underbrace{
    \sum_{j\in\N_i} \sin( \varphi_{\k,i} - \varphi_{\k,j} )
  }_{=0} \notag \\
& \phantom{= \omega_0} - \varepsilon \frac{K}{2d} \sum_{j\in\N_i} \cos( \varphi_{\k,i} - \varphi_{\k,j} ) \,\Delta_{\l,i}  \notag \\
& \phantom{= \omega_0} + \varepsilon \frac{K}{2d} \sum_{j\in\N_i} \cos( \varphi_{\k,i} - \varphi_{\k,j} ) \,\Delta_{\l,j}
  + \mathcal{O}(\varepsilon^2) \notag \\
&\approx \omega_0 - \varepsilon \Delta_{\l,i} \frac{K}{2d} \sum_{j\in\N_i} \cos( \k\cdot\x_{ij} ) [ 1 - \exp( -i\,\l\cdot\x_{ij} ) ]
  \quad. \notag
\end{align}
}
\begin{equation}
\label{eq:lambda_l}
\lambda_{\k\l} = -\frac{K T_0}{2d} \sum_{j\in\N_i} \left[ 1 - \cos(\l\cdot\x_{ij}) \right] \cos(\k\cdot\x_{ij}) \quad,
\end{equation}
where $\x_{ij} = \x_j-\x_i$ such that $|\x_{ij}|=a$ for $j\in\N_i$.
Hence,
periodic solutions with $|\k|<\pi/(2a)$ are linearly stable,
while periodic solutions with $|\k|>\pi/(2a)$ can be saddle nodes or linearly unstable.

Let $N_i = \max_j N_j$ be the number of oscillators
along the longest direction of the $N_1{\times}\ldots{\times}N_d$-unit cell.
The slowest decaying perturbation mode is then
$\l_\mathrm{max} = 2\pi/L\,\e_i$,
where we introduce system length $L=N_i\,a$.
For the Lyapunov exponent of the slowest decaying perturbation mode $\l$ of the dominant wave solution $\k=\zero$,
we thus find, analogous to the one-dimensional case treated above
\begin{align}
\max_{\l\in\K\setminus\{\zero\}} \lambda_{\zero\l} 
&= -\frac{K T_0}{d} 
  \left[ 1 - \cos\left(\frac{2\pi a}{L}\right) \right] \notag \\
&\approx 
-\frac{K T_0}{d} 
  \left(\frac{2\pi a}{L}\right)^2 
\sim L^{-2}
\end{align}
to leading order in $a/L$,
where $L=a \max\{N_1,\ldots,N_d\}$ denotes system length.
This maximal Lyapunov exponent sets a relaxation time
of the dominant wave solution, 
$\tau_\mathrm{relax} = T_0 \left| \max \lambda_{\zero\l} \right|^{-1}$.

\subsection{Relation to XY model}

One can map the Kuramoto model with identical phase oscillators and sinusoidal coupling, Eq.~\eqref{eq:Kuramoto_sine},
to an equilibrium system by switching to a co-rotating frame 
with variables $\theta_i = \varphi_i - \omega_0 t$.
Specifically, we consider the Hamilton of the classical XY model
\begin{equation}
\label{eq:H}
H = -J \sum_{j\in\N_i} \cos(\theta_i - \theta_j)
\end{equation}
and consider the over-damped dynamics
\begin{equation}
\label{eq:theta}
\gamma\,\dot{\theta}_i = - \frac{\partial}{\partial \theta_i} H \quad.
\end{equation} 
Here, $\gamma$ denotes an effective friction coefficient.
Eq.~\eqref{eq:theta} is equivalent to Eq.~\eqref{eq:Kuramoto_sine} for
\begin{equation}
J = \frac{\gamma K}{4d} \quad.
\end{equation}
Fixed points $\boldsymbol{\theta}_\k^\ast$ of Eq.~\eqref{eq:theta} [over-damped XY model]
correspond exactly to periodic solutions 
$\FP_\k(t)=\omega_0t\, \mathbf{1}+\boldsymbol{\theta}_\k^\ast$ 
of Eq.~\eqref{eq:Kuramoto_sine} [Kuramoto model with local coupling].
For small perturbations $\varepsilon\boldsymbol{\Delta}$ from a stable fixed point $\boldsymbol{\theta}_\k^\ast$,
we can approximate the Hamiltonian $H$ as a harmonic potential
\begin{equation}
H(\boldsymbol{\theta}_\k^\ast + \varepsilon \boldsymbol{\Delta} ) 
\approx H(\boldsymbol{\theta}_\k^\ast) + \frac{\varepsilon^2}{2} \, \Delta H \quad,
\end{equation}
where 
$
\Delta H =
\boldsymbol{\Delta} \cdot 
\nabla^2 H_{|\boldsymbol{\theta}=\boldsymbol{\theta^\ast_\k}} \cdot 
\boldsymbol{\Delta}^\dagger
$ 
and $\dagger$ denotes the complex conjugate of a transposed vector.
We can interpret $\Delta H$ either an effective spring stiffness along the direction of the perturbation $\boldsymbol{\Delta}$, 
or as a normalized energy penalty of the perturbation mode $\boldsymbol{\Delta}$.
We have a direct relationship between the Lyapunov exponents $\lambda_{\zero\l}$ of the Kuramoto model
for the dominant wave solution $\k=\zero$, 
as given in Eq.~\eqref{eq:lambda_l},
and the energy penalties 
$\Delta H_\l = \Delta H(\boldsymbol{\Delta}_\l)$
of the fundamental perturbation modes $\boldsymbol{\Delta}_\l$ defined in Eq.~\eqref{eq:delta_l}.
A short calculation shows
\footnote{
Specifically,
\begin{align}
\Delta H_\l 
&= 
\boldsymbol{\Delta_\l} \cdot
\nabla^2 H_{|\boldsymbol{\theta}=\boldsymbol{\theta^\ast_\k}} \cdot 
\boldsymbol{\Delta_\l}^\dagger 
\notag \\
&=
-J\, \sum_{r,s}
\frac{\partial}{\partial \theta_r} \frac{\partial}{\partial \theta_s} \sum_{i,j\in\N_i} \cos(\theta_i-\theta_j) \,
\Delta_{\l,r} \, \Delta_{\l,s}^\ast 
\notag \\
&= 2J \!\! \sum_{i,j\in\N_i}
 \cos(\k\cdot\x_{ij}) \, [ \Delta_{\l,i}\,\Delta_{\l,i}^\ast - \Delta_{\l,j}\,\Delta_{\l,i}^\ast ]
\notag \\
&= \frac{\gamma K}{2d} \, N \!\! \sum_{j\in\N_i}
 \cos(\k\cdot\x_{ij}) \, [ 1 - \exp( - i\,\l\cdot\x_{ij}) ]
\quad. \notag
\end{align}
}
\begin{equation}
\lambda_{\zero\l} \,/\, T_0 =
- \frac{1}{\gamma} \,
  \frac{\Delta H_\l}{N} 
\quad.
\end{equation}
Here, $T_0=2\pi/\omega_0$ is the period of the periodic solutions.

The Hamiltonian $H$ possesses $O(2)$-symmetry;
any spontaneous ``magnetization'' with $|\langle e^{i\theta_j}\rangle|>0$ corresponds to spontaneous symmetry breaking.
For $d\ge 3$ space dimensions (i.e., $\Lambda\subset\R^d$), 
the classical XY model is known to exhibit a conventional phase transition with spontaneous magnetization below a critical temperature $T_c$.
For $d=2$ dimensions, there is no long-range order at any finite temperature, 
and thus no conventional phase transition.
This is a consequence of the famous Mermin-Wagner theorem that rules out long-range order
in two-dimensional systems with local coupling and continuous symmetries \cite{Mermin1966}.
In these systems, the energy penalty for long-wavelength perturbations of the ordered ground state is independent of system size;
hence these Goldstone bosons become thermally excited at any finite temperature.
Nonetheless, for $d=2$, the classical XY model exhibits a so-called Kosterlitz-Thouless transition,
from a disordered high-temperature state with exponential decay of spatial correlations,
to a quasi-ordered low-temperature state with algebraic decay of spatial correlations \cite{Chaikin1995},
at a critical temperature $k_B T_c/J \approx 0.89$ \cite{Mattis1984}.
\end{document}